\documentclass[10pt,conference]{IEEEtran}
\IEEEoverridecommandlockouts
\usepackage{cite}
\usepackage{booktabs}
\usepackage{multirow}
\usepackage{amsmath,amssymb,amsfonts}
\usepackage{algorithmic}
\usepackage{graphicx}
\usepackage{textcomp}
\usepackage{xcolor}
\usepackage{subcaption}
\usepackage{tcolorbox}
\usepackage{algorithm}
\usepackage{enumitem}
\usepackage{tabularx}
\usepackage{amssymb}
\usepackage{url}
\usepackage{hyperref}
\def\BibTeX{{\rm B\kern-.05em{\sc i\kern-.025em b}\kern-.08em
    T\kern-.1667em\lower.7ex\hbox{E}\kern-.125emX}}
\begin{document}


\title{TrajAudit: Automated Failure Diagnosis for Agentic Coding Systems}

\author{\IEEEauthorblockN{1\textsuperscript{st} Minxing Wang}
\IEEEauthorblockA{
\textit{Singapore Management University}\\
Singapore, Singapore \\
mx.wang.2026@phdcs.smu.edu.sg}
\and
\IEEEauthorblockN{2\textsuperscript{nd} Xiaofei Xie}
\IEEEauthorblockA{
\textit{Singapore Management University}\\
Singapore, Singapore \\
xfxie@smu.edu.sg}
\and
\IEEEauthorblockN{3\textsuperscript{rd} Yintong Huo}
\IEEEauthorblockA{
\textit{Singapore Management University}\\
Singapore, Singapore \\
ythuo@smu.edu.sg}
}

\maketitle

\begin{abstract}
Agentic systems have been widely studied to automate coding tasks such as bug fixing and feature implementation. As these systems increasingly operate on complex codebases, understanding where and why they fail becomes essential for iterative refinement and operational reliability. Existing automated failure diagnosis approaches leverage \textit{task execution trajectories}, yet they struggle with trajectories produced by repository-level coding agents due to two key properties. First, these trajectories are often long, spanning many execution steps, making it difficult for LLMs to track the causal chain of failure over the execution history. Second, these trajectories are laden with noise, containing substantial low-signal observations such as redundant program structures and verbose code context, which can interfere with LLM reasoning.

To address these challenges, we propose \textit{TrajAudit}, an automated failure diagnosis framework specifically for trajectories produced by repository-level coding agents. TrajAudit employs an investigator agent supported by two modules: one reduces failure-irrelevant noisy context through semantic saliency folding, and the other derives preliminary diagnostic guidance from test failure reports as prior knowledge to help LLMs focus on likely failure regions. The investigator agent can further invoke tools to inspect folded content on demand, enabling a focused investigation without losing access to the full trajectory context. We also introduce \textit{RootSE}, a benchmark of 102 real-world agentic failure instances from repository-level coding tasks, each annotated with the earliest decisive error step and a diagnostic justification. Experiments on RootSE show that TrajAudit outperforms the strongest baselines by 10.8\% and 21.6\% in exact failure localization accuracy in the with- and without-reference settings, respectively, demonstrating its effectiveness.
\end{abstract}


\section{Introduction}

LLM-based agentic systems are autonomous systems powered by large language models (LLMs) that perceive environmental states, perform goal-oriented reasoning, and execute actions in a closed-loop manner~\cite{albrecht2018autonomous, franklin1996agent, luo2025large}.  Recently, a growing number of LLM-based agents have been developed to automate \textit{repository-level coding tasks}, like issue resolution and feature development~\cite{hong2023metagpt, zhang2024autocoderover, wang2024openhands, hu2025compileagent, qian2024chatdev}.

While these agents succeed in simple tasks such as single-file modifications, they still struggle with complex, multi-file tasks that require long-horizon reasoning and execution~\cite{han2024llm, epperson2025interactive, xia2023automated, liurepobench}. These failures occur in opaque ways, often as the cumulative consequence of a single early mistake, such as a misunderstanding of the task requirements or a flawed implementation plan\cite{epperson2025interactive, parnin2011automated}. Therefore, understanding where and why agents fail is critical for the iterative refinement of agentic systems and, ultimately, trustworthy intelligent software engineering~\cite{parnin2011automated, barrak2025traceability, hou2024large, lu2025exploring}.

\begin{figure*}[t] 
  \centering
  \includegraphics[width=0.78\textwidth]{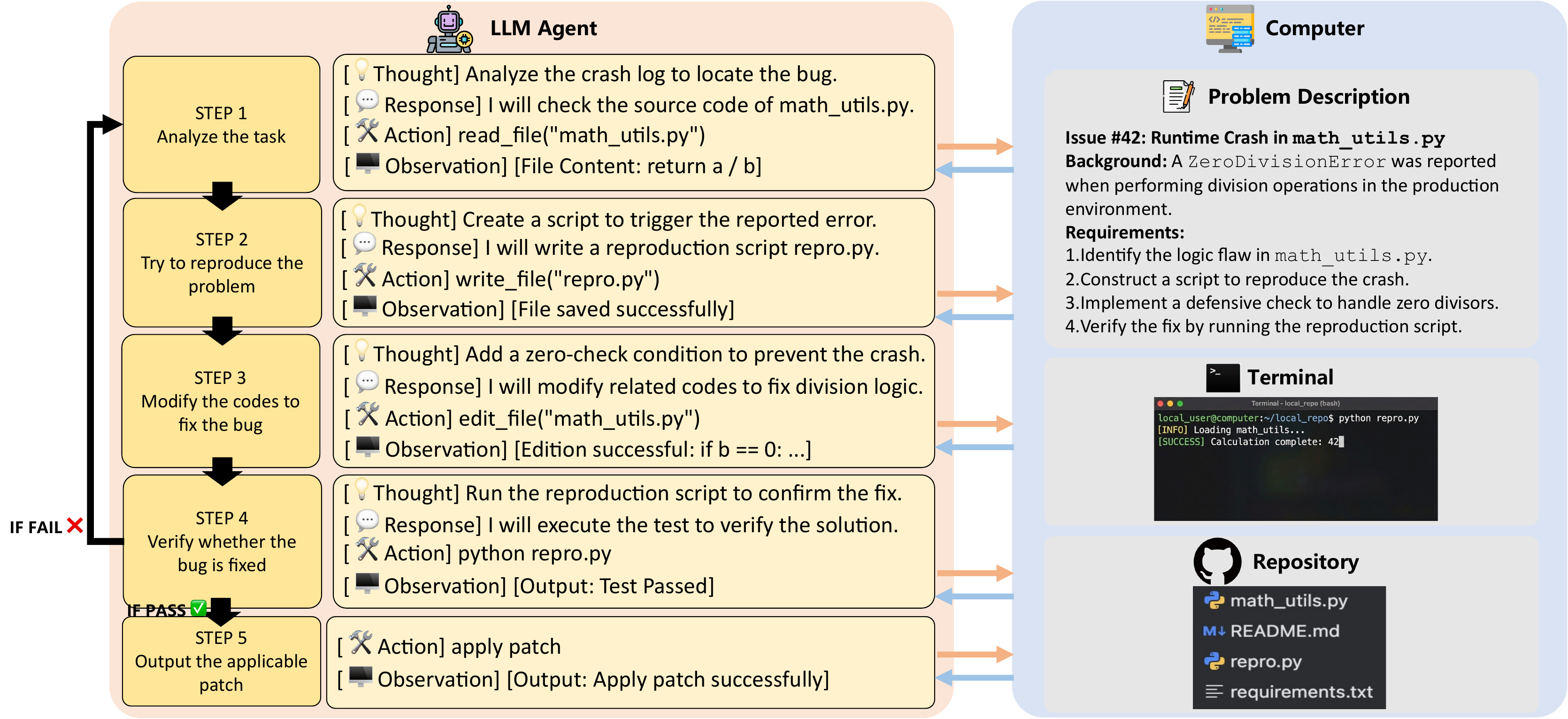}
  \caption{The agent workflow and execution trajectory in a coding task.}
  \label{fig:se_agentic_framework}
\end{figure*}

\textit{Execution trajectory}, which records sequential steps of the agent's reasoning, tool invocations, and environmental observations, provides the key information to monitor the agent's behavior~\cite{pan2025multiagent, zhang2025agentracer}. As shown in Figure~\ref{fig:se_agentic_framework}, each step contains four types of information: the thinking process (Thought), the transitional language produced by the agent (Response), the actions taken (Action), and the environmental feedback (Observation). These trajectories are commonly used for understanding failures behind agent execution.
For example, one pioneering study conducted in-depth manual investigations of inter-step inconsistencies to characterize failure patterns, such as reasoning-action conflicts~\cite{bouzenia2025understanding}. To further automate \textit{failure diagnosis}, existing LLM-based approaches ask LLMs to inspect complete trajectories~\cite{zhang2025which, deshpande2025trail}, while spectrum-based approaches identify suspicious steps by contrasting failed and successful runs~\cite{ge2025introducing}. However, neither handles long and noisy repository-level coding trajectories well.


We identify two distinct challenges as follows,
\textit{(1) Observational noise.}
Observations refer to the information returned by tools invoked by the agent, often accounting for over 70\% of the total trajectory content~\cite{zhang2025which}. However, most observations are not directly relevant to diagnosis, such as redundant program structures and verbose code context, which would complicate automated failure diagnosis \cite{shi2023large}.
\textit{(2) Excessive length.} These trajectories often span from 20 to over 100 steps, with each step containing on average over 5,000 characters. Existing studies \cite{chaudhury2025epman, tian2025selective, liu2024lost} show that even state-of-the-art LLMs struggle to maintain reasoning quality when processing long contexts.

Both challenges stem from a fundamental limitation of existing diagnosis methods: they passively consume entire trajectories, treating all steps as equally relevant. Inspired by the action-taking nature of agents \cite{qin2023toolllm, schick2023toolformer}, we propose an agentic approach that \textit{actively explores and fetches} the most relevant fragments from long and noisy trajectories.

To this end, we introduce \textbf{TrajAudit}, an automated failure diagnosis framework for repository-level coding agent trajectories, including pinpointing the first step at which the agent commits a decisive error (i.e., \textit{earliest decisive error step}) and providing diagnosis justifications.
TrajAudit addresses the aforementioned challenges through an agent with two synergistic modules. 
(1) \textit{Prior failure reasoning} derives a preliminary diagnosis by prompting an LLM to identify the most suspicious region responsible for the failure based on the test case errors. This diagnosis is then incorporated into the investigator agent's context as prior knowledge, directing the model's focus on failure segments, thereby mitigating long-context degradation.
(2) \textit{Semantic saliency folding} selectively compresses trajectory observations by retaining only failure-relevant context, such as code patch structures and entries containing failure indicators (e.g., 'fail', 'exception').
(3) \textit{Investigator agent} enables dynamic access to the full trajectory, performing on-demand retrieval of folded content through predefined interactive tools. This mechanism enables a top-down diagnostic approach, allowing the diagnosis to begin with a high-level overview and selectively drill down into details on demand, thereby mitigating observational noise.

Furthermore, to evaluate TrajAudit, we introduce \textbf{RootSE}, the first benchmark to evaluate a model's ability to diagnose agent execution failure in completing repository-level coding tasks. The dataset consists of 102 instances with over 5,000 execution steps, offering a challenging testbed for failure diagnosis. Experimental results on RootSE show that TrajAudit consistently outperforms all baselines, achieving up to 21.6\% improvement in exact step-level accuracy over the strongest baseline, demonstrating its effectiveness for diagnosing coding agents.


To sum up, the main contributions are threefold:
\begin{itemize}
    \item We introduce TrajAudit, an automated failure diagnosis framework for repository-level coding agent trajectories, which employs an investigator agent supported by semantic saliency folding and prior failure reasoning to localize the earliest decisive error step and generate diagnostic justifications.
    \item We construct RootSE, the first benchmark for earliest-step 
    failure diagnosis on repository-level coding task trajectories, 
    consisting of 102 real-world instances annotated with 
    earliest decisive error steps and diagnostic justifications.
    \item We release all artifacts to facilitate independent verification, reproducibility, and future research.
\end{itemize}
\section{Preliminaries}
\label{sec:preliminaries}
In this section, we formally define the execution trajectory and the failure diagnosis task, which serve as the foundation for the benchmark and methods in later sections.

\subsection{Execution Trajectory}
We model an LLM-based agentic system as a stateful system that interacts with a software environment through a sequence of discrete steps to complete a given coding task. Following prior work~\cite{zhang2025which}, we represent a full execution trajectory as:
\begin{equation}
\tau = (s_{0}, \eta_{1}, s_{1}, \eta_{2}, s_{2}, \ldots, \eta_{T}, s_{T})
\end{equation}
where $s_{0}$ is the initial system state, $s_{t}$ is the system \textit{state} after step $t$, and $\eta_{t}$ denotes the agent-environment \textit{interaction step} at time $t$. A \textit{state} represents the available execution context, including the task description, repository snapshot, accumulated interaction history, and tool outputs. Each \textit{interaction step} is represented by four components: a \textbf{Thought} (the reasoning trace explicitly logged by the agent), a \textbf{Response} (the transitional natural language statement generated by the agent to describe its current or intended operation), an \textbf{Action} (the tool invocation by the agent), and an \textbf{Observation} (the environmental feedback returned by the tools). We define a binary outcome function $Z(\tau) \in \{0, 1\}$, where $Z(\tau) = 1$ indicates task failure and $Z(\tau) = 0$ indicates task success. In the context of coding tasks, task success is determined by whether all acceptance test cases pass.

\subsection{Failure Diagnosis}
The failure diagnosis task is defined as follows: given a task description, a failed trajectory, the test code, and the corresponding test failure output, identify the first step at which the agent commits a decisive error (i.e., the earliest decisive error step) and provide a natural language justification explaining how that step leads to task failure.

We adopt the definition of the earliest decisive error step from Zhang et al.~\cite{zhang2025which}. Given a failed trajectory $\tau$ with $Z(\tau) = 1$, we define $\eta_{t}$ as a decisive error step if there exists an alternative interaction $\bar{\eta}_{t}$ such that replacing $\eta_{t}$ with $\bar{\eta}_{t}$, while keeping all preceding steps unchanged, yields a modified trajectory $\bar\tau$ satisfying $Z(\bar\tau) = 0$. Among all decisive error steps in $\tau$, the one with the smallest step index is defined as the earliest decisive error step. 
\section{Related Work}
\label{sec:related_work}

\subsection{Failure Diagnosis Methods}

\subsubsection{LLM-based} 
Zhang et al.~\cite{zhang2025which} proposed three methods that feed trajectories into LLMs through single-pass, sequential, and binary-search strategies to identify the decisive failure step. Deshpande et al.~\cite{deshpande2025trail} introduced a taxonomy of 20 agentic error types, such as formatting errors and instruction non-compliance, and located all failed steps within a trajectory by prompting an LLM with both the taxonomy and the full trajectory. 
These methods can be effective for tasks with short execution traces, such as web browsing. However, their effectiveness collapses on coding task trajectories. Such long trajectories may exceed the context window of LLMs, making methods that require the full trajectory impractical. In addition, tool outputs in coding task trajectories often contain substantial noise, such as irrelevant code context and verbose file structures. Such noise interferes with the LLM's ability to identify the decisive failure step.

\subsubsection{Spectrum-based}
Ge et al.~\cite{ge2025introducing} proposed FAMAS, which adapts the idea of spectrum-based fault localization to diagnose agentic systems. FAMAS first re-executes the failed task multiple times to gather a set of trajectories. It then normalizes each into standardized behavior triples using an LLM and clusters them into semantically equivalent groups. Finally, it ranks groups by frequency-based suspiciousness score to pinpoint failure indicators.

While FAMAS outperforms prior purely LLM-based methods on general-purpose task failure localization~\cite{ge2025introducing}, it faces two limitations when applied to coding tasks. First, repeated re-execution is costly, as each run costs nearly one dollar on average~\cite{jimenez2023swe}, making scaling debugging impractical. Second, by representing agent behavior through action-based triples, FAMAS overlooks reasoning traces, which are a key failure signal in coding agents.

\subsubsection{Learning-based}
Zhang et al.~\cite{zhang2025agentracer} proposed AgenTracer, which fine-tunes a lightweight model on annotated failure trajectories via reinforcement learning to identify the decisive failure step. However, this approach requires training a dedicated model, making it less flexible to deploy in new settings.

\subsection{Failure Diagnosis Benchmarks for Agentic Systems}
Zhang et al.~\cite{zhang2025which} proposed Who\&When, the first benchmark for failure diagnosis in agentic systems, comprising 184 annotated trajectories with ground-truth labels for the earliest decisive error step. However, their trajectories came from general-purpose tasks, which tend to be shorter and less noisy than coding ones, limiting the applicability to complex SE settings. Deshpande et al.~\cite{deshpande2025trail} proposed TRAIL, a benchmark of 148 annotated trajectories, of which 118 are information retrieval tasks and only 30 involve software engineering. 
Furthermore, TRAIL recognizes all plausible erroneous steps, such as formatting errors and hallucinations. In contrast, our work focuses on identifying the earliest crucial step that leads to the failure.
\section{The RootSE Benchmark}
\label{sec:rootse}
\begin{figure}[t]
    \centering
    \begin{subfigure}[b]{0.9\linewidth}
        \includegraphics[width=\linewidth]{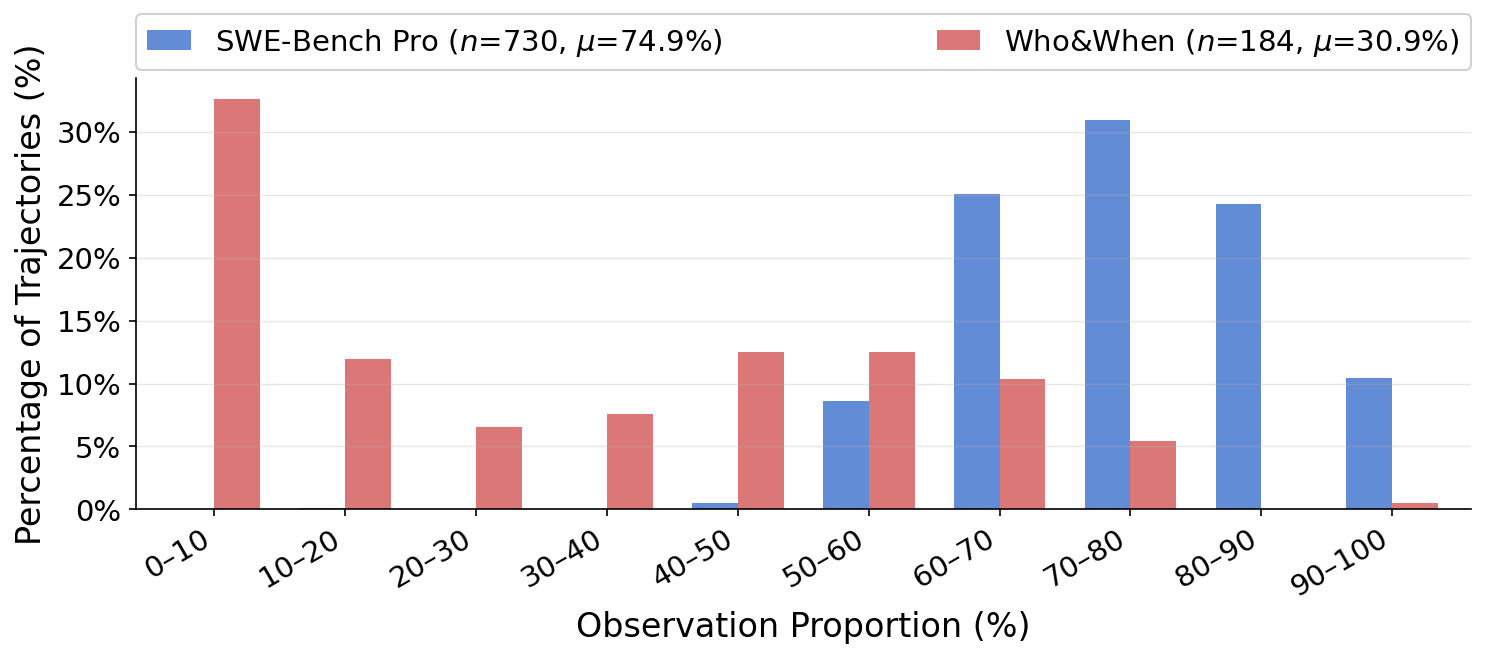}
        \caption{Observation proportion.}
        \label{fig:obs-pct}
    \end{subfigure}
    \hfill
    \begin{subfigure}[b]{0.9\linewidth}
        \includegraphics[width=\linewidth]{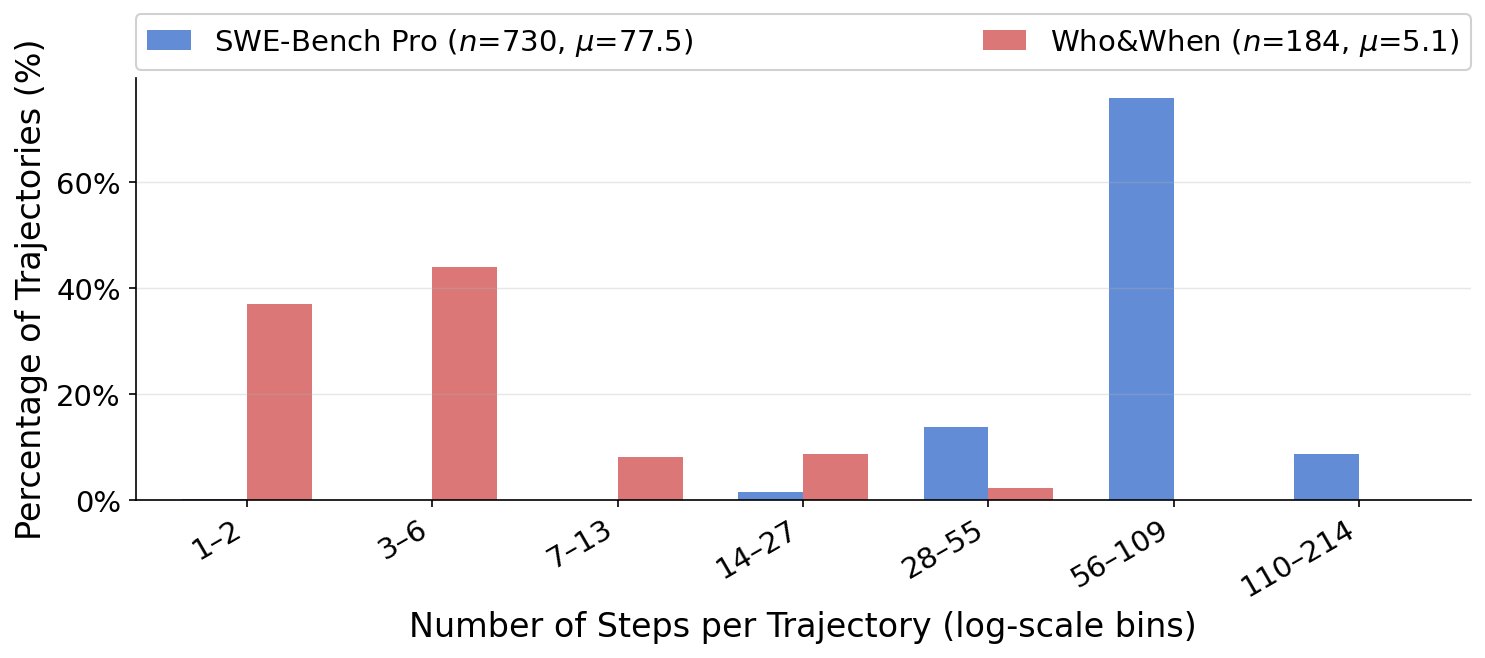}
        \caption{Steps per trajectory.}
        \label{fig:traj-length}
    \end{subfigure}
    \caption{Comparison of trajectory characteristics between SWE-bench Pro and Who\&When.}
    \label{fig:traj-compare}
\end{figure}

Who\&When, the most closely related benchmark for earliest step failure diagnosis, consists of trajectories from general-purpose tasks such as web browsing~\cite{mialon2023gaia}. As shown in Figure~\ref{fig:traj-compare}, trajectories from general-purpose tasks differ from those produced during repository-level coding tasks (e.g., SWE-bench Pro) in two important ways: (1) observation content accounts for a substantially smaller proportion of the trajectory, and (2) the trajectories span far fewer steps~\cite{deng2025swe}. This lower proportion of observation content suggests that reasoning, actions, and their outcomes are less obscured by noisy context such as irrelevant code snippets and repository file structures, reducing the amount of low-signal context that LLMs must process. These differences suggest that Who\&When does not fully capture the difficulty of coding tasks, and no existing benchmark evaluates the earliest step failure diagnosis in this setting.

To fill this gap, we introduce RootSE, the first benchmark for earliest step failure diagnosis on repository-level coding task trajectories, characterized by long-horizon execution as well as substantial observation content. It comprises 102 failed instances generated by four representative agents tackling diverse repository-level coding problems, each annotated with the earliest decisive error step and a diagnosis justification. Together, these instances encompass more than 5,000 individual execution steps and approximately 30 million characters, representing a challenging and realistic evaluation setting.

\subsection{Task Description}
\label{subsec:task_description}
Each instance in RootSE comprises four core elements:
\textbf{(1) Task Specification}: encompasses all metadata required for task execution and completion verification, including the task description, repository name, base commit, and test code patch; \textbf{(2) System Configuration}: identifies the specific agentic system and the underlying LLM employed; \textbf{(3) Failure Context}: consists of the complete execution trajectory and the test error messages; and \textbf{(4) Ground-truth Labels}: includes the earliest error step, diagnosis justification, and the reference patch for the task.

In particular, RootSE asks a diagnosis model to take the task description, failure trajectory, test code, and corresponding error messages as input, and output the failure step, along with its justification.
RootSE employs three metrics to evaluate the results: \textbf{(1) Exact Step-Level Accuracy}, which measures the percentage of instances where the predicted step matches the ground truth exactly; \textbf{(2) Tolerated Step-Level Accuracy}, which represents the proportion of predictions that fall within a predefined tolerance window around the ground truth; and \textbf{(3) Justification Accuracy}, which measures the percentage of instances where the predicted diagnosis justification is semantically equivalent to the ground truth, as checked by LLM-as-a-judge. 

\subsection{Data Collection}

\subsubsection{Agentic System Selection}
We select four agentic systems as trajectory sources: SWE-agent, Live-SWE-agent, OpenHands, and AutoCodeRover. These systems are either widely adopted or technically distinctive. SWE-agent introduces the Agent-Computer Interface (ACI), which constrains the action space and standardizes environment feedback to improve the reliability of agent-environment interactions~\cite{yang2024swe}. Live-SWE-agent builds upon SWE-agent with a single prompt modification that enables the agent to create task-specific tools on demand~\cite{xia2025live}. OpenHands provides a sandboxed runtime environment along with a web-based GUI that supports iterative software maintenance workflows~\cite{wang2024openhands}. SWE-agent and OpenHands have each received over 15,000 GitHub stars. AutoCodeRover represents programs as abstract syntax trees, enabling code search at the granularity of classes and methods for efficient task resolution~\cite{zhang2024autocoderover}, and has been commercialized.

We restrict trajectory sources to single-agent systems for two reasons. First, RootSE focuses on diagnosing failures caused by an agent's own reasoning and actions, such as misunderstanding task requirements and incomplete code implementations. Since research indicates that 37\% of multi-agent system (MAS) failures stem from inter-agent cooperation breakdowns~\cite{pan2025multiagent}, including MAS trajectories would introduce confounding factors that fall outside the scope of this benchmark. Second, since each agent in a MAS operates on its own individual trajectory, TrajAudit can still be applied to diagnose reasoning and action failures of individual agents within MAS settings.

\subsubsection{Task Selection}

We select task sources based on three criteria: tasks must originate from real-world repository-level issues, support automatic verification through executable test suites, and require multi-step code reasoning and execution. Based on these criteria, we adopt SWE-bench~\cite{jimenez2023swe}, SWE-bench Pro~\cite{deng2025swe}, and SWE-rebench~\cite{badertdinov2026swe} as task sources. SWE-bench comprises 2,294 problems from real GitHub issues across 12 Python repositories. SWE-bench Pro targets more challenging cross-file tasks, requiring modifications spanning an average of 4.1 files across multiple programming languages. SWE-rebench is an automatically constructed dataset of over 21,000 tasks designed to provide unseen problems to mitigate the contamination risk.

\subsubsection{Trajectory Generation}
We collected candidate trajectories by running the selected agentic systems on the chosen benchmarks, using representative LLM backbones including ChatGPT, Qwen, Claude, and Gemini series, and by incorporating trajectories from the public dataset~\cite{trofimova2025openhandstrajs}, yielding an initial pool of 500 candidate trajectories. These trajectories were then filtered through manual inspection in three steps. First, we removed trajectories with ambiguous task descriptions, leaving 412 trajectories.
Second, we excluded cases whose test suites checked requirements not mentioned in the task description, leaving 364 trajectories.
Third, we retained only cases with exactly one failing test, since trajectories with multiple failing tests may reflect simultaneous deviations from several distinct requirements, making it difficult to identify a single earliest decisive error step, leaving 102 trajectories. The resulting benchmark spans 35 repositories, three programming languages, and four agent architectures; a detailed analysis is provided in Section~\ref{subsec:complexity_analysis}.

\subsection{Annotation Procedure}
After collecting data, we annotate each instance with the earliest decisive step index and a corresponding justification. To ensure annotation quality, the procedure involves three stages and three human experts: two annotators ($A_{1}, A_{2}$), each with three years of software development experience and prior experience using coding agents, and one validator ($V_{0}$) with extensive experience in developing coding agents, who arbitrates unresolved disagreements between $A_{1}$ and $A_{2}$.

\textbf{Stage $\mathrm{\textbf{I}}$:} In the first stage, the three experts independently review the task definition and draft candidate annotation criteria. They then consolidate these criteria into a unified guideline through group discussion, as shown in Figure~\ref{fig:annotation_guideline}. \textbf{Stage $\mathrm{\textbf{II}}$:} In the second stage, $A_{1}$ and $A_{2}$ independently annotate the entire dataset following the unified guideline. 
Each annotation consists of two core elements: the index of the earliest decisive error step and a textual justification explaining how this step causes the system to deviate from the correct solution and ultimately leads to task failure. 
\textbf{Stage $\mathrm{\textbf{III}}$:} In the final stage, $A_{1}$ and $A_{2}$ discuss all annotation inconsistencies to reach a consensus. For any unresolved disagreements, $V_{0}$ first independently annotates the disputed cases and then joins the discussion to reach a final consensus.

\begin{figure}[t]
    \centering
    \begin{tcolorbox}[colback=yellow!10!white, colframe=yellow!50!black, title=\textbf{Annotation Guideline}, sharp corners, fontupper=\footnotesize]
\textbf{Key Notes:}\\
a) An error can occur at the reasoning level (flawed planning or incorrect inference) or the execution level (sound reasoning but faulty action, such as improper tool use or faulty code editing).

b) A step is marked as an error only when the agent has settled on a flawed direction, not when it explores multiple hypotheses, even if some of them are incorrect.

\textbf{Justification:} \\Provide an explanation of why this step is the earliest decisive error step and how it leads to the eventual task failure.

\textbf{Example:} \\At Step 1, the agent misreads the issue as requiring a new function rather than a bug fix, causing the entire implementation to address the wrong objective and ultimately failing all test cases.
\end{tcolorbox}
    \caption{RootSE annotation guideline.}
    \label{fig:annotation_guideline}
\end{figure}

To evaluate annotation reliability, we measure the inter-rater agreement between $A_{1}$ and $A_{2}$ following Stage $\mathrm{II}$. The Cohen's Kappa coefficient for earliest decisive error step identification is 0.78, indicating substantial agreement~\cite{landis1977measurement, cohen1960coefficient, 10.1016/j.infsof.2008.09.009}. All remaining discrepancies are subsequently resolved in Stage $\mathrm{III}$ through the arbitration of $V_{0}$, producing the final consensus annotations used in RootSE. To further validate the annotations, we conduct a counterfactual verification study on a random sample of 30 instances. For each instance, the annotators construct a corrective replacement for the annotated step without consulting the reference patch, for example, replacing flawed reasoning with a corrected diagnosis or plan, or replacing a faulty code edit with a corrected one, while keeping the preceding trajectory unchanged. The agent is then resumed from the modified trajectory under the same execution setting. All 30 instances result in successful task completion. Although this result does not prove that the annotated step is the unique cause of failure, it provides empirical support that it lies on a causal path to the observed failure.

\subsection{Benchmark Analysis}
We analyze RootSE from two perspectives: failure phase distribution and the complexity and diversity of tasks and trajectories.

\subsubsection{Failure Phase Distribution Analysis}
\begin{figure}[t]
    \centering
    \includegraphics[width=0.85\linewidth]{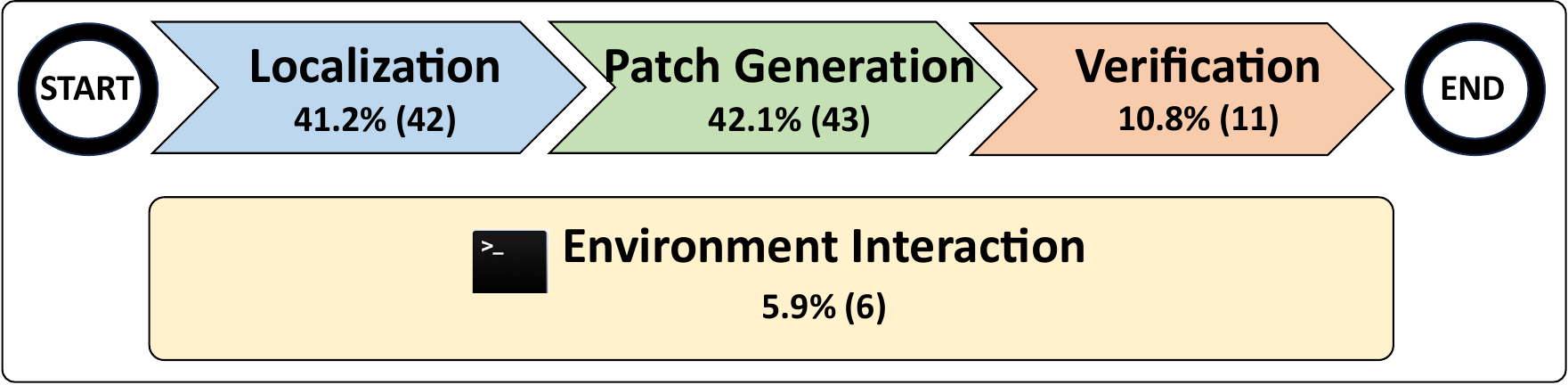}
    \caption{Phase-wise failure distribution in RootSE.}
    \label{fig:failure_modes}
\end{figure}

Following the standard workflow of automated program repair, which consists of the Localization Phase, Patch Generation Phase, and Verification Phase~\cite{10.1145/3631974}, we use these phases as a coarse-grained abstraction of repository-level code modification. Based on the purpose of the annotated error step in each trajectory, we map each RootSE instance to the corresponding phase or category. As shown in Figure~\ref{fig:failure_modes}, 
failures are not confined to a single phase, with the Localization Phase accounting for 41.2\% (42 instances), the Patch Generation Phase for 42.1\% (43 instances), and the Verification Phase for 10.8\% (11 instances), indicating that RootSE captures failure scenarios across the main activities of locating, modifying, and validating code changes. In addition, we identify environment-interaction failures as an orthogonal category (5.9\%, 6 instances), including failures in tool usage or command execution. These failures may arise at different points in the whole coding process. 

\subsubsection{Complexity and Diversity Analysis}
\label{subsec:complexity_analysis}
We evaluate RootSE's complexity along two dimensions: task complexity and trajectory complexity. Table~\ref{tab:dataset_comparison} compares statistics between RootSE and Who\&When dataset.

\begin{table}[tbp]
\footnotesize
    \centering
    \caption{Comparison of dataset complexity across multiple dimensions for RootSE and Who\&When.}
    \label{tab:dataset_comparison}
    \begin{tabular}{lcc}
        \toprule
        \textbf{Metric} & \textbf{Who\&When} & \textbf{RootSE (Ours)} \\
        \midrule
        \textbf{Problem Source} & Personal Assistant & Software Maintenance \\
        \textbf{\#Task Desc. Chars} & 240.5 & \textbf{8,186.7} \\
        \textbf{\#Steps} & 5.2 & \textbf{51.7} \\
        \textbf{\#Chars per Step} & 3,919.0 & \textbf{5,567.8} \\
        \bottomrule
    \end{tabular}
\end{table}

For task complexity, we use one metric: the average character count of task descriptions (\#Task Desc. Chars). As shown in Table~\ref{tab:dataset_comparison}, task descriptions in RootSE average 8,186.7 characters, about 34 times longer than the 240.5 characters in Who\&When, reflecting substantially higher contextual complexity.

For trajectory complexity, we adopt two metrics: the number of steps per trajectory (\#steps) and the average character count per step (\#chars per step). As shown in Table~\ref{tab:dataset_comparison}, on average, RootSE trajectories contain nearly ten times as many steps as Who\&When, and each step is also 42\% longer. These extended and highly detailed trajectories impose higher requirements on failure diagnosis methods.

Beyond complexity, RootSE exhibits substantial diversity across repositories, languages, agent architectures, and task types. The 102 instances span 35 distinct repositories across three programming languages: Python (73 instances, 71.6\%), Go (24 instances, 23.5\%), and JavaScript (5 instances, 4.9\%). Trajectories are sourced from four structurally diverse agentic systems: SWE-agent (53), OpenHands (30), AutoCodeRover (10), and Live-SWE-agent (9). Tasks are predominantly bug fixing, with approximately 15\% involving feature implementation. This diversity across repositories, languages, and agent architectures ensures that RootSE captures a broad range of real-world failure scenarios rather than artifacts of any single system or domain.
\section{TrajAudit Methodology}

\subsection{Overview} 
\begin{figure}[t]
    \centering
    \includegraphics[width=0.48\textwidth]{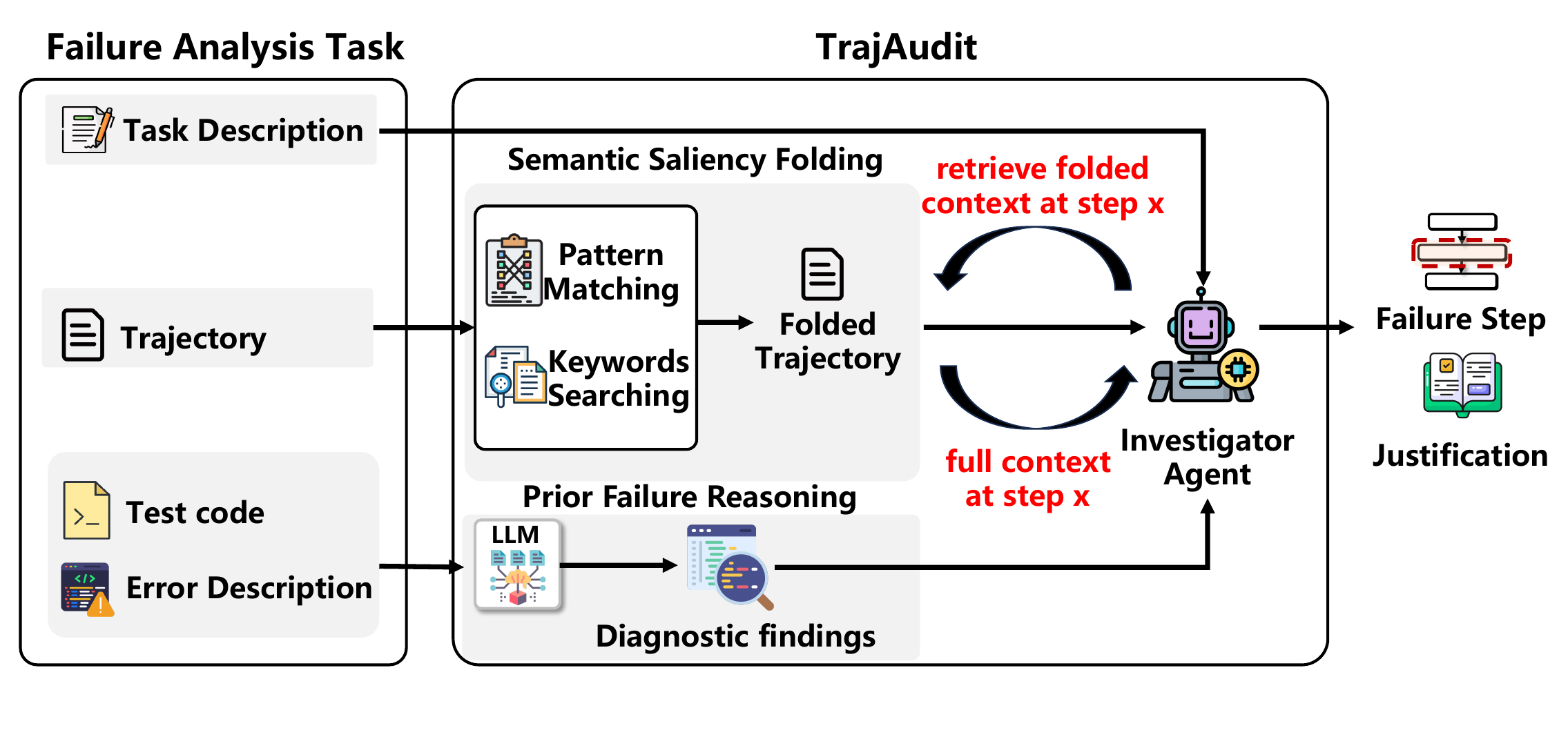}
    \caption{The overall workflow of TrajAudit.}
    \label{fig:trajaudit_overview}
\end{figure}

TrajAudit comprises an investigator agent and two supporting modules: a prior failure reasoning (PFR) module and a semantic saliency folding (SSF) module. As illustrated in Figure~\ref{fig:trajaudit_overview}, SSF compresses failure-irrelevant observations into placeholders, reducing observational noise while preserving critical signals such as code patches and error indicators. PFR generates a preliminary diagnosis from the test code and error description, directing the investigator agent toward the most probable failure region to mitigate long-context degradation~\cite{wei2022chain, weiser1984program}. Serving as a central hub, the investigator agent integrates both outputs and dynamically retrieves folded content on demand~\cite{schick2023toolformer}, enabling a focused top-down investigation without sacrificing access to the full trajectory. We detail each component in the following subsections.

\subsection{Prior Failure Reasoning}
The prior failure reasoning module aims to provide prior knowledge to guide the LLM toward the most relevant areas of the trajectory for diagnosis.
This design is inspired by the debugging process of software engineers. Rather than analyzing an entire execution trajectory at once, engineers typically look at the error information (e.g., exception logs) to localize the potential failure region before expanding their scope of investigation. Although recent LLMs have substantially extended context windows, existing research demonstrates that long context reasoning still incurs significant performance degradation~\cite{zhou2025gsm}. We therefore adopt a similar strategy to help the agent maintain focus when encountering long-horizon trajectories.

Specifically, this module accepts the failure information from task execution, including test code and error description, and prompts the LLM to identify the workflow phase most likely responsible for the failure along with the corresponding rationale, without constraining it to a predefined set of phases. This preliminary diagnostic serves as prior knowledge, directing the LLM's attention toward the most probable failure region before it processes the full trajectory.

\subsection{Semantic Saliency Folding}
Compared to natural language reasoning or web browsing tasks, coding trajectories are much denser and more complex, particularly within the observation steps that document code artifacts and execution states.
In RootSE, for example, observation content accounts for over 74.9\% of the total trajectory length. While certain observations contain critical failure signals (e.g., error logs or patch outputs), the majority constitute redundant data. For instance, creating a file often returns verbose outputs such as complete directory trees. All such tool responses are logged entirely within the observation entries of the trajectory. As these failure-irrelevant observations accumulate over long-horizon tasks, they introduce substantial noise that degrades the reasoning capabilities of diagnostic methods.

To mitigate this, we introduce the semantic saliency folding module, which identifies potential failure signals through pattern and keyword matching and replaces the remaining irrelevant observations with recoverable folding markers. We consider two types of observations as failure-relevant. The first encompasses the generated code patches, as they explicitly capture how the agent modifies the code and thus allow us to identify which exact code changes lead to the test failure. Second, inspired by run-time monitoring for traditional systems \cite{10.1109/TSE.2014.2372785, du2017deeplog, guo2021logbert, he2016experience, landauer2023deep}, we preserve the observations containing failure-indicative keywords, such as \texttt{traceback} and \texttt{exception}.

The semantic saliency folding is detailed in Algorithm \ref{alg:fold_algorithm}. To begin with, we apply pattern matching to identify code patches. The widely adopted patch format in software projects is the code diff \cite{myers1986nd}, which follows a unified structure: the patch header contains file metadata in the form of \texttt{{-}{-}{-} a} and \texttt{+++ b}, representing the original and modified files respectively, while the hunk header follows the format \texttt{@@ -N,M +P,Q @@} to indicate the location of modifications within the code file. We encode these fixed structural patterns into regular expressions and apply them to identify patches within observations. Afterwards, we predefine a failure indicator dictionary covering the frequently occurring failure-indicative keywords, constructed through LLM generation and manual refinement. Any observation containing these keywords is flagged as a signal that the agentic system has likely encountered an anomalous state. Observations that lack both patch data and failure keywords are folded, while those matching either criterion are preserved. This folding process is reversible: when the compressed trajectory lacks sufficient evidence, the investigator agent can retrieve the original content of any folded observation through the inspection tool.

The resulting folded trajectory is then passed to the investigator agent for further analysis.

\begin{algorithm}[t]
\footnotesize
\renewcommand{\algorithmicrequire}{\textbf{Input:}}
\renewcommand{\algorithmicensure}{\textbf{Output:}}
\renewcommand{\algorithmiccomment}[1]{$\triangleright$ #1}
\caption{Semantic Saliency Folding}
\label{alg:fold_algorithm}
\begin{algorithmic}
\REQUIRE Trajectory $T$
\ENSURE Folded trajectory $T^{*}$

\STATE $P \leftarrow \text{compiled patch pattern}$ \COMMENT{e.g., \texttt{{-}{-}{-}a}, \texttt{+++b}, \texttt{@@ -N,M +P,Q @@}}

\STATE $K \leftarrow \{\text{"exception", "fail", "error", "traceback", "invalid" \ldots}\}$

\STATE $T^{*} \leftarrow []$

\FOR {each entry $c$ in $T$}
    \STATE $d \leftarrow parse(c)$
    \IF{$d[\text{"observation"}] \not\sim P \textbf{ AND } \nexists k \in K: k \in d[\text{"observation"}]$}
    \STATE $d[\text{"observation"}] \leftarrow \text{"Folded, invoke API to view the full context"}$
    \ENDIF
    \STATE $\text{append } serialized(d) \text{ to } T^{*}$
\ENDFOR

\end{algorithmic}
\end{algorithm}

\subsection{Investigator Agent}
The investigator agent serves as a central hub that integrates the processed outputs from the above two modules, and dynamically probes the folded observations on demand to make the final diagnosis. 

The investigator agent is prompted to first assess whether the current context is sufficient for failure diagnosis. If so, it directly outputs the identified failure step index along with the corresponding justification; otherwise, it iteratively invokes local tools to inspect folded observations and progressively expands the available context until a final diagnosis can be produced.

To support this process, the agent is equipped with two interactive tools: one for retrieving the content of a folded observation at a specific step, and one for submitting the final diagnostic result.
\section{Experimental Setup}
We evaluate TrajAudit by answering the following research questions (RQs):
\begin{itemize}[leftmargin=10pt, itemsep=0.5em]
    \item \textbf{RQ1:} How effective is TrajAudit?
    \item \textbf{RQ2:} How does TrajAudit balance localization effectiveness and token cost compared with baselines?
    \item \textbf{RQ3:} How robust is TrajAudit across different backbones?
    \item \textbf{RQ4:} How does each component contribute to TrajAudit?
\end{itemize}

\subsection{Dataset}
We conduct experiments on RootSE, the dataset proposed in this paper (Section~\ref{sec:rootse}), comprising 102 instances collected from trajectories of representative agentic systems that failed to resolve real-world repository-level coding issues. Each instance is annotated with the earliest decisive error step and the corresponding diagnosis justification.

\subsection{Baselines}
We compare TrajAudit against five baselines described in Section~\ref{sec:related_work}: three strategies proposed by Zhang et al.~\cite{zhang2025which} (All-at-Once, Step-by-Step, Binary Search), TRAIL~\cite{deshpande2025trail}, and FAMAS~\cite{ge2025introducing}. We exclude AgenTracer~\cite{zhang2025agentracer} as its trained model parameters are not publicly available.
To adapt these methods to RootSE, we embed the RootSE annotation guideline into every baseline's prompt, ensuring all methods operate under the same definition of the earliest decisive error step.
For TRAIL, we adopt a two-step pipeline. In the first step, we follow the original design by feeding the complete trajectory along with TRAIL's error taxonomy to the LLM to identify all failed steps. In the second step, we provide all the identified failed steps together with the annotation guideline to the LLM, which selects the single earliest decisive error step.
For FAMAS, we provide at least three trajectories per case for suspiciousness scoring, prioritizing successful trajectories when available. These trajectories are obtained from multiple runs of the same agent system on the same task. For cases where no successful trajectory exists, we supplement with additional failed trajectories from different runs. The resulting suspiciousness rankings are then combined with the annotation guideline and fed to the LLM to produce the final prediction.
For All-at-Once, Step-by-Step, and Binary Search, we incorporate the annotation guidelines into the prompts. Cases where the trajectory exceeds the LLM's context limit are treated as diagnosis failures.

\subsection{Evaluation Settings}
Inspired by the evaluation protocol of Who\&When, we conduct experiments under two settings: \textbf{with reference} and \textbf{without reference}, depending on whether a reference patch for the current task is provided. Unlike Who\&When, where tasks have deterministic answers, repository-level coding tasks support multiple valid solutions, making a single definitive solution patch infeasible. We therefore provide a reference patch as one representative solution.

These two settings evaluate failure diagnosis methods under different real-world scenarios. The with-reference setting reflects the common development scenario where agentic systems are evaluated on benchmark tasks with known reference solutions, allowing diagnosis methods to leverage reference information when identifying failure causes. In contrast, the without-reference setting reflects practical deployment scenarios where no reference solution is available and diagnosis must rely solely on execution logs.

\subsection{Metrics}
We evaluate all methods using three metrics introduced in Section~\ref{subsec:task_description}. To validate the reliability of the LLM-as-a-judge approach used for Justification Accuracy, we randomly sampled 30 instances per method (180 instances in total) and had two annotators independently assess whether each predicted justification is semantically equivalent to the ground truth. The two annotators reached an initial agreement rate of 97.2\%, and all remaining disagreements were resolved through discussion to produce a consensus label. The LLM-as-a-judge achieved an agreement rate of 93.3\% with the consensus labels, indicating sufficient reliability as an automated metric. For short, we denote step-level accuracy under tolerance $k$ as Tol@$k$. Unless otherwise specified, we use Tol@$0$ under the without-reference setting as the primary metric. 
 
\subsection{Implementation Details}
We employ Claude Sonnet 4.5\footnote{\texttt{claude-sonnet-4-5-20250929}} as the default backbone LLM. To reduce randomness and improve reproducibility, the temperature is set to 0.
\section{Evaluation}

\subsection{RQ1. How effective is TrajAudit?}
We comprehensively evaluate the effectiveness of TrajAudit against state-of-the-art baselines across four dimensions: exact step-level accuracy, tolerated step-level accuracy, performance across varying trajectory lengths, and justification accuracy.

\subsubsection{Exact Step-level Accuracy}
As shown in Table~\ref{tab:exact_accuracy}, TRAIL and FAMAS achieve relatively low exact step-level accuracy on RootSE, despite their competitive performance on general-purpose agentic tasks such as web browsing and file understanding. This suggests that these methods do not generalize well to repository-level coding task trajectories. In contrast, TrajAudit achieves the highest exact step-level accuracy in both settings. It outperforms the best-performing baseline TRAIL by 10.8 percentage points (49.0\% vs. 38.2\%) with reference and by 21.6 percentage points (52.0\% vs. 30.4\%) without reference.

\begin{table}[!htbp]
\centering
\caption{Exact step-level accuracy (\%) on RootSE.}
\label{tab:exact_accuracy}
\begin{tabular}{llcc}
\toprule
\textbf{Category} & \textbf{Method} & \textbf{w/ Reference} & \textbf{w/o Reference} \\
\midrule
\multirow{4}{*}{LLM-based} 
    & All-at-Once   & 27.5          & 24.5          \\
    & Step-by-Step  & 23.5          & 22.6          \\
    & Binary Search & 18.6          & 11.8          \\
    & TRAIL         & \underline{38.2} & \underline{30.4} \\
\midrule
Spectrum-based 
    & FAMAS         & 25.5          & \underline{30.4} \\
\midrule
Agent-based 
    & \textbf{TrajAudit} & \textbf{49.0} & \textbf{52.0} \\
\bottomrule
\end{tabular}
\vspace{3pt}
\begin{minipage}{\linewidth}
\footnotesize
$^\dagger$ Cases exceeding the LLM context limit are treated as diagnosis failures (All-at-Once: 3, Binary Search: 2, TRAIL: 1).
\end{minipage}
\end{table}

To confirm statistical significance, we apply exact McNemar's test on per-instance paired results under the without-reference setting, where $b$ and $c$ denote the number of instances correctly diagnosed only by TrajAudit and only by the baseline, respectively. As shown in Table~\ref{tab:mcnemar}, TrajAudit significantly outperforms all five baselines (p $<$ 0.001 in each case), with substantially larger $b$ than $c$ in every comparison. These results indicate that TrajAudit's advantage is consistent across instances and is not attributable to chance.

\begin{table}[tbp]
\centering
\caption{McNemar's test results comparing TrajAudit against each baseline (Tol@0, without reference). $b$ = instances correctly diagnosed only by TrajAudit; $c$ = instances correctly diagnosed only by the baseline.}
\label{tab:mcnemar}
\begin{tabular}{llccc}
\toprule
\textbf{Category} & \textbf{Baseline} & $b$ & $c$ & $p$-value \\
\midrule
\multirow{4}{*}{LLM-based}
 & All-at-Once  & 34 & 6  & $<0.001$ \\
 & Step-by-Step & 42 & 12 & $<0.001$ \\
 & Binary-Search & 44 & 3 & $<0.001$ \\
 & TRAIL        & 29 & 7  & $<0.001$ \\
\midrule
Spectrum-based & FAMAS & 28 & 6 & $<0.001$ \\
\bottomrule
\end{tabular}
\end{table}

We further observe that the relative effectiveness of the with- and without-reference settings varies across methods. For methods that process the full trajectory in a single pass (All-at-Once, Step-by-Step, Binary Search, and TRAIL), the with-reference setting consistently yields higher accuracy. We attribute this to the reference patch providing an anchor for these methods: by identifying the intended code changes, the LLM can narrow its focus to relevant trajectory regions despite the long and noisy context. In contrast, TrajAudit and FAMAS perform better in the without-reference setting. Both methods employ mechanisms that already provide focused step-level views of the trajectory, namely active on-demand inspection and suspiciousness-based clustering respectively, reducing their reliance on external anchoring. In this case, the reference patch may instead introduce bias by directing the LLM toward patch-related code changes, causing it to overlook earlier decisive errors in planning or exploration phases.

\subsubsection{Tolerated Step-level Accuracy}
In many practical scenarios, pinpointing the exact failure step is not always necessary; localizing the failure within a narrow range of candidate steps is often sufficient for downstream debugging and correction. Table~\ref{tab:tolerance} reports step-level accuracy under different tolerance windows in the without-reference setting. As the tolerance increases, all methods improve, yet TrajAudit consistently achieves the highest accuracy across all tolerance levels. At $\pm$3, TrajAudit reaches 65.7\%, outperforming TRAIL and FAMAS by 15.7 and 13.7 percentage points, respectively. This consistent gap suggests that TrajAudit not only localizes failure steps more precisely, but also maintains its advantage under relaxed evaluation criteria.

\begin{table}[tbp]
\centering
\caption{Step-level accuracy under different tolerances (without-reference setting).}
\label{tab:tolerance}
\begin{tabular}{llcccc}
\toprule
\textbf{Category} & \textbf{Method} & Tol@$0$ & Tol@$1$ & Tol@$2$ & Tol@$3$ \\
\midrule
\multirow{4}{*}{LLM-based}
    & All-at-Once   & 24.5             & 41.2             & 50.0             & \underline{52.0} \\
    & Step-by-Step  & 22.6             & 34.3             & 41.2             & 44.1             \\
    & Binary Search & 11.8             & 27.5             & 34.3             & 39.2             \\
    & TRAIL         & \underline{30.4} & 42.2             & 49.0             & 50.0             \\
\midrule
Spectrum-based
    & FAMAS         & \underline{30.4} & \underline{44.1} & \underline{51.0} & \underline{52.0} \\
\midrule
Agent-based
    & \textbf{TrajAudit} & \textbf{52.0} & \textbf{58.8} & \textbf{62.8} & \textbf{65.7} \\
\bottomrule
\end{tabular}
\end{table}

\subsubsection{Performance across Varying Trajectory Lengths}
\begin{figure}[!htbp]
    \centering
    \includegraphics[width=0.9\linewidth]{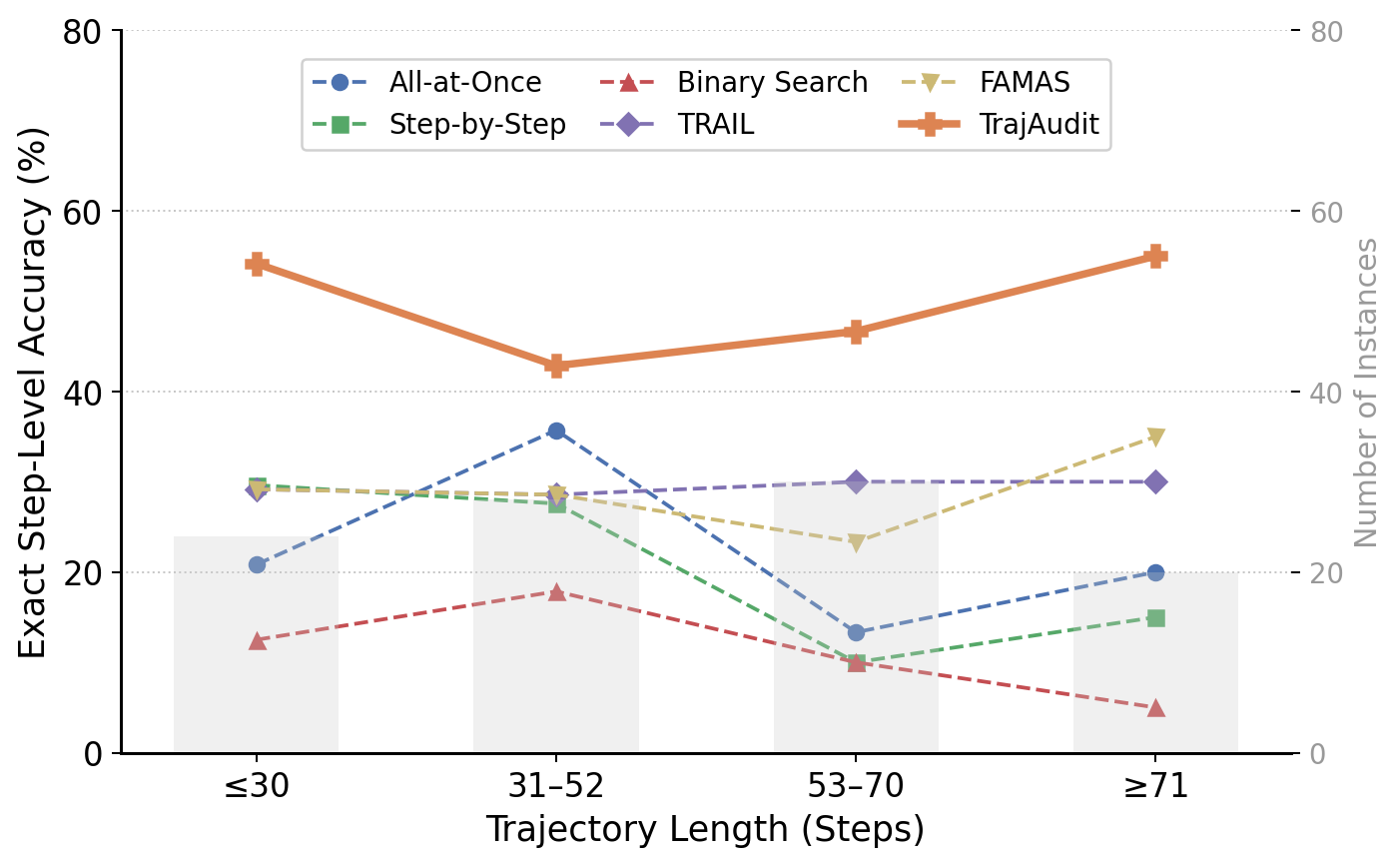}
    \caption{Exact step-level accuracy across varying trajectory length on RootSE.}
    \label{fig:complexity_level}
\end{figure}

Figure~\ref{fig:complexity_level} reports exact step-level accuracy across different trajectory-length buckets in the without-reference setting. TrajAudit achieves the highest accuracy in every bucket, indicating that its advantage is not limited to trajectories of a specific length. For example, TrajAudit reaches 46.7\% accuracy on trajectories with 53--70 steps and 55.0\% accuracy on trajectories with at least 71 steps, while the best-performing baseline achieves 30.0\% and 35.0\% in the same buckets, respectively.

\subsubsection{Justification Accuracy}
Beyond step-level localization accuracy, we further evaluate whether each method produces a justification that correctly identifies the annotated failure cause. Table~\ref{tab:justification} reports the justification accuracy of each method, assessed by \texttt{Claude Sonnet 4.5} as the LLM evaluator using the same evaluation rubric for all methods. Binary Search is excluded from this comparison, as it outputs only a region label (upper or lower half) at each iteration and produces no diagnostic justification. TrajAudit achieves the highest justification accuracy of 52.9\%, outperforming the best baseline (All-at-Once, 42.4\%) by 10.5 percentage points, while Step-by-Step yields only 14.7\%, suggesting that its fragmented per-step analysis may struggle to capture the broader trajectory context required for accurate failure attribution.

\begin{table}[tbp]
\centering
\caption{Justification accuracy (\%) between predicted and ground-truth justifications in the without-reference setting.}
\label{tab:justification}
\resizebox{\linewidth}{!}{%
\begin{tabular}{lccccc}
\toprule
 & All-at-Once & Step-by-Step & TRAIL & FAMAS & TrajAudit \\
\midrule
Accuracy (\%) & \underline{42.4} & 14.7 & 40.6 & 28.4 & \textbf{52.9} \\
\bottomrule
\end{tabular}}
\end{table}

\subsection{RQ2. How does TrajAudit balance localization effectiveness and token cost compared with baselines?}

\begin{figure}[!htbp]
    \centering
    \includegraphics[width=0.8\linewidth]{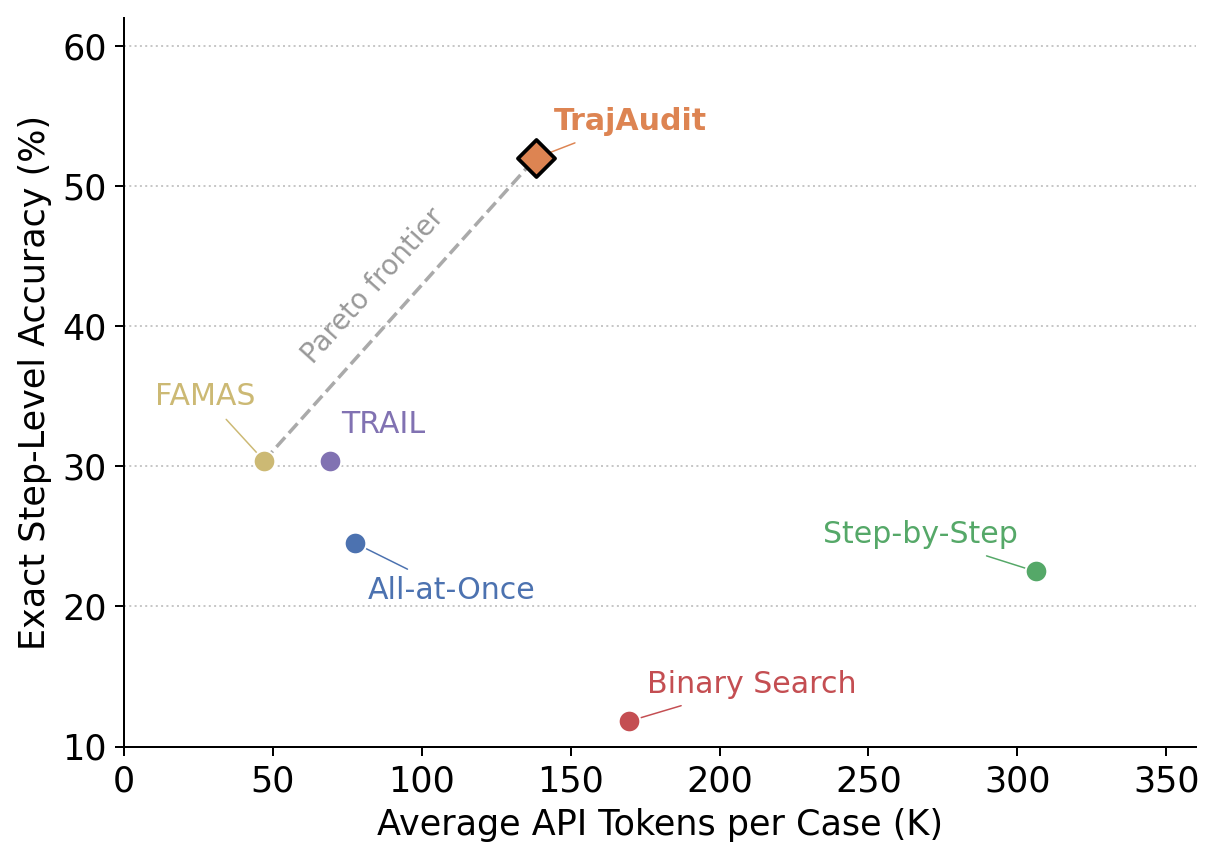}
    \caption{Effectiveness--cost trade-off under the without-reference
    setting (Tolerance=0). The dashed line connects
    the observed non-dominated methods for visualization only.}
    \label{fig:cost_accuracy}
\end{figure}

Figure~\ref{fig:cost_accuracy} compares all methods under the
without-reference setting in terms of exact step-level accuracy and average API token consumption per case, where the latter is computed as total tokens consumed divided by the 102 benchmark cases. TrajAudit achieves the highest accuracy of 52.0\% at an average cost of 138K tokens per case. It strictly dominates Binary Search (169K, 11.8\%) and Step-by-Step (306K, 22.6\%), achieving higher accuracy while consuming fewer tokens. Compared with the lower-cost baseline FAMAS (47K, 30.4\%), TrajAudit requires more tokens but improves accuracy by 21.6 percentage points. The remaining baselines are dominated by either FAMAS or TrajAudit. Together with FAMAS, TrajAudit lies on the observed Pareto frontier among the evaluated methods, i.e., neither method is dominated by another evaluated method in both accuracy and token cost. This places TrajAudit at the high-accuracy end of the frontier and indicates a favorable balance between localization effectiveness and computational cost when accuracy is prioritized.

\subsection{RQ3. How robust is TrajAudit across different backbones?}

Table~\ref{tab:rq3-backbone} reports TrajAudit's performance with five different backbone LLMs. Across all backbones, TrajAudit achieves 40.2--55.9\% Tol@0 and 61.8--70.6\% Tol@3. Even with the weakest evaluated backbone, TrajAudit achieves a higher Tol@0 than the best-performing baseline (\textsc{Trail}, 30.4\%) reported in RQ1 under the same without-reference setting, supporting that its improvement is not specific to a single backbone choice. However, the 15.7 percentage-point Tol@0 gap and the 20.5 percentage-point justification gap between the strongest and weakest backbones suggest that backbone choice still affects both localization and explanation quality.

The results further reveal that localization accuracy and justification accuracy are not perfectly aligned. Gemini 3.1 Pro achieves the best localization performance (55.9\% Tol@0, 70.6\% Tol@3), yet its justification accuracy (45.9\%) is lower than that of Claude Sonnet 4.5 and GPT-5.2. Conversely, Claude Sonnet 4.5 obtains the highest justification accuracy of 52.9\%, despite ranking second in localization accuracy. These results suggest that identifying the faulty step and articulating a semantically accurate explanation require related but distinct capabilities. Overall, TrajAudit remains applicable across diverse backbone LLMs, while backbone choice continues to influence both localization accuracy and explanation quality.

\begin{table}[tbp]
\centering
\caption{TrajAudit performance with different backbone LLMs on RootSE.}
\label{tab:rq3-backbone}
\setlength{\tabcolsep}{6pt}
\resizebox{\columnwidth}{!}{
\begin{tabular}{lrrrrr}
\toprule
\textbf{Backbone LLM} & \textbf{Tol@0} & \textbf{Tol@1} & \textbf{Tol@2} & \textbf{Tol@3} & \textbf{Justif.} \\
\midrule
\raisebox{-0.1\height}{\includegraphics[height=0.9em]{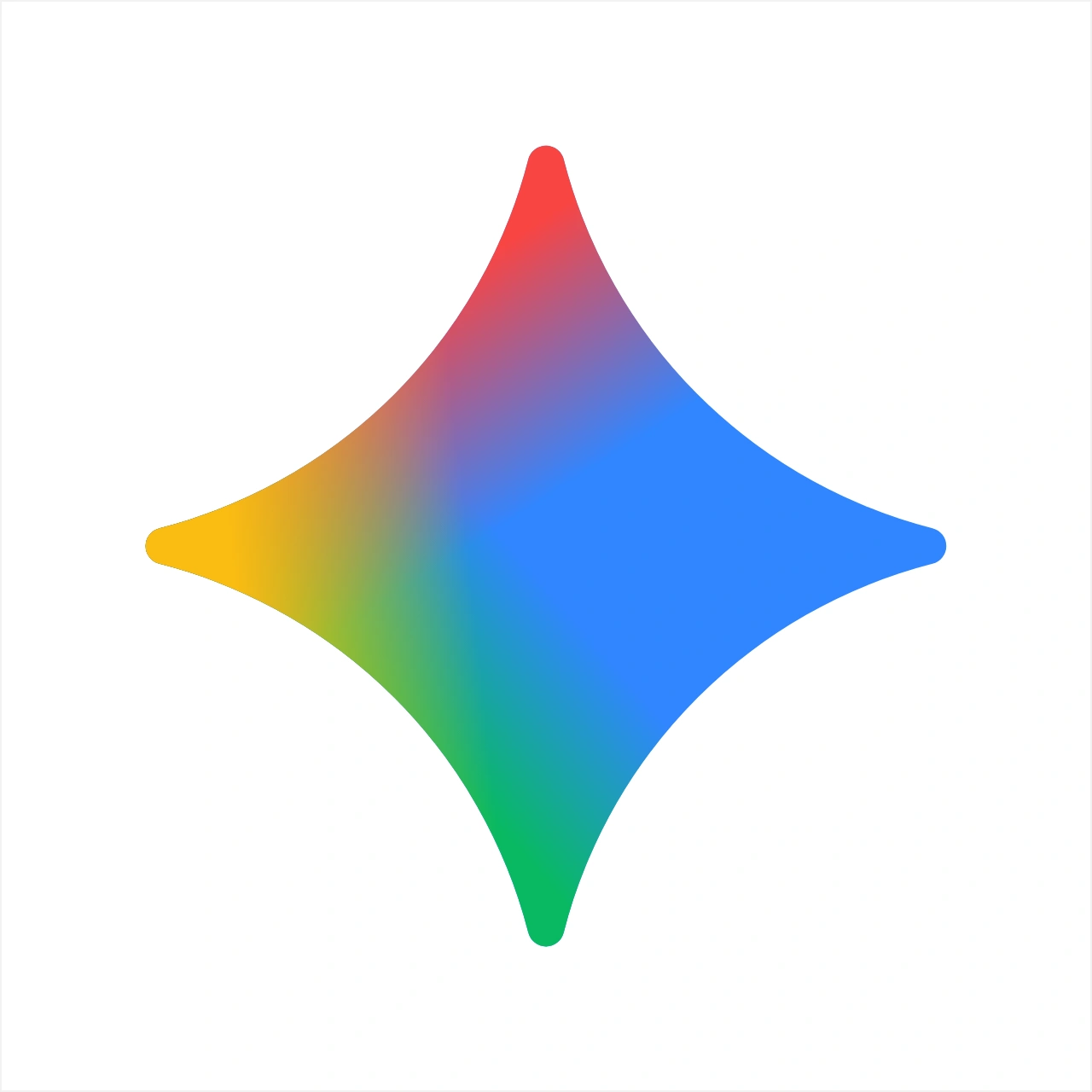}}~\textsc{Gemini 3.1 Pro}
  & \textbf{55.9} & \textbf{65.7} & \textbf{67.7} & \textbf{70.6} & 45.9 \\
\raisebox{-0.1\height}{\includegraphics[height=0.9em]{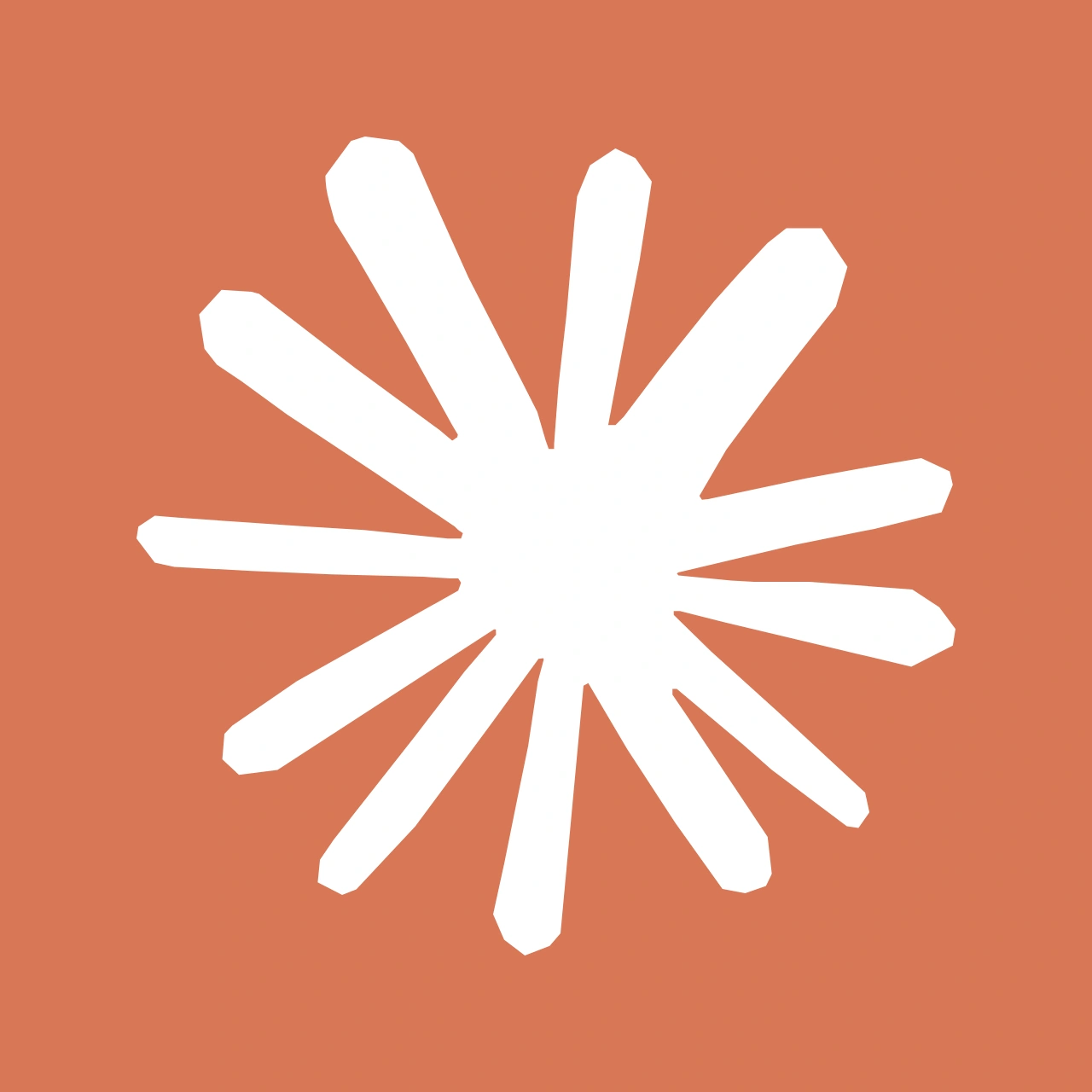}}~\textsc{Claude Sonnet 4.5}
  & 52.0 & 58.8 & 62.8 & 65.7 & \textbf{52.9} \\
\raisebox{-0.1\height}{\includegraphics[height=0.9em]{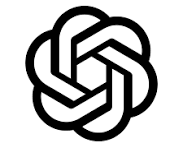}}~\textsc{GPT-5.2}
  & 45.1 & 54.9 & 59.8 & 63.7 & 52.0 \\
\raisebox{-0.1\height}{\includegraphics[height=0.9em]{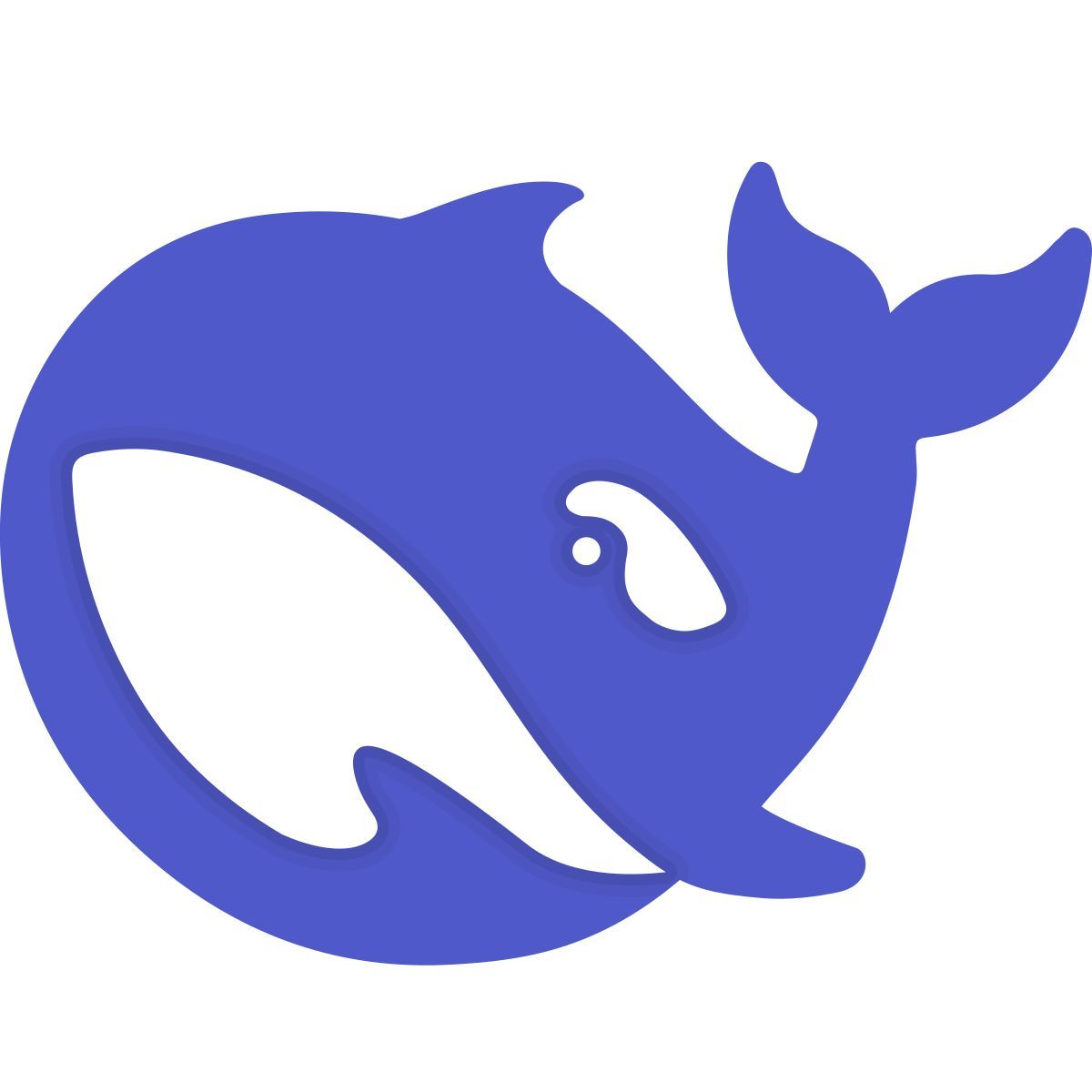}}~\textsc{DeepSeek V4 Pro}
  & 45.1 & 57.8 & 61.8 & 64.7 & 42.6 \\
\raisebox{-0.1\height}{\includegraphics[height=0.9em]{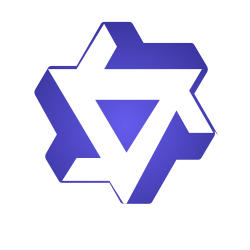}}~\textsc{Qwen3-235B}
  & 40.2 & 52.9 & 58.8 & 61.8 & 32.4 \\
\bottomrule
\end{tabular}
}
\end{table}

\subsection{RQ4. How does each component contribute to TrajAudit?}

\begin{table}[!htbp]
\caption{Ablation study of TrajAudit components in the without-reference setting.}
\label{tab:combined_ablation}
\centering
\small
\setlength{\tabcolsep}{3pt}
\begin{tabularx}{\columnwidth}{ccXXX}
\toprule
\textbf{SSF} & \textbf{PFR} & \textbf{Exact Acc. (\%)} & \textbf{Just. Acc. (\%)} & \textbf{Avg. Tokens (K)} \\
\midrule
\checkmark & \checkmark & \textbf{52.0} & \textbf{52.9} & 138.2 \\
\checkmark & $\times$   & 34.3 ($\downarrow$17.7) & 38.0 ($\downarrow$14.9) & 125.5 ($\downarrow$12.7) \\
$\times$   & \checkmark & 38.2 ($\downarrow$13.8) & 36.0 ($\downarrow$16.9) & 86.2 ($\downarrow$52.0) \\
\bottomrule
\addlinespace[1ex]
\multicolumn{5}{l}{\footnotesize Numbers in parentheses denote absolute changes from the full model;} \\
\multicolumn{5}{l}{\footnotesize accuracy in percentage points, tokens in K.} \\
\multicolumn{5}{l}{\footnotesize SSF: Semantic Saliency Folding; PFR: Prior Failure Reasoning.}
\end{tabularx}
\end{table}

Table~\ref{tab:combined_ablation} presents the ablation results for each component of TrajAudit in the without-reference setting. Removing PFR causes the largest drop in exact step-level accuracy, decreasing it by 17.7\% while reducing token consumption by only 12.7K tokens per case, suggesting that PFR provides a strong localization prior with relatively small token overhead. Removing SSF reduces token consumption by 52.0K tokens per case but also
decreases exact accuracy by 13.8\% and justification accuracy by 16.9\%. Although SSF folds the majority of trajectory steps upfront, TrajAudit's iterative multi-step localization process accumulates tokens across successive rounds of on-demand expansion calls, resulting in higher overall consumption than the variant without SSF. Overall, this indicates that SSF's token overhead is accompanied by substantial gains in both localization and justification quality.

To further understand SSF's behavior, we analyze its folding operations across all 102 benchmark cases. On average, SSF folds the observations of 94.6\% of trajectory steps per case, retaining the full observations of only 5.4\% of steps, primarily those containing patch content or failure-related signals such as exceptions and tracebacks.
Among the folded steps, the agent subsequently expands only 20.2\% on demand via targeted tool calls, leaving over 75\% of all trajectory steps never expanded in full. These results suggest that SSF enables a focused investigation strategy: it aggressively abstracts low-saliency steps while allowing the agent to selectively retrieve specific context when needed.

\section{TrajAudit Error Analysis}

To understand the limitations of TrajAudit, we analyze the cases where TrajAudit fails to localize the annotated earliest decisive error step. We find that the localization accuracy correlates with the position of the annotated error step. When the error occurs within the first 20\% of steps, accuracy drops to 19\%--26\% across backbones, whereas errors in the middle portion (35\%--65\%) achieve substantially higher accuracy of 54\%--57\%. Early errors are difficult to localize because they leave no observable transition from correct to incorrect behavior~\cite{318773.318946}. The entire subsequent trajectory executes on a flawed basis, with no clear turning point to identify. When the error occurs in the middle of the trajectory, by contrast, TrajAudit can draw on the contrast between the preceding correct steps and the subsequent erroneous ones to localize the mistake more reliably.

\begin{figure}[!htbp]
    \centering
    \includegraphics[width=0.85\linewidth]{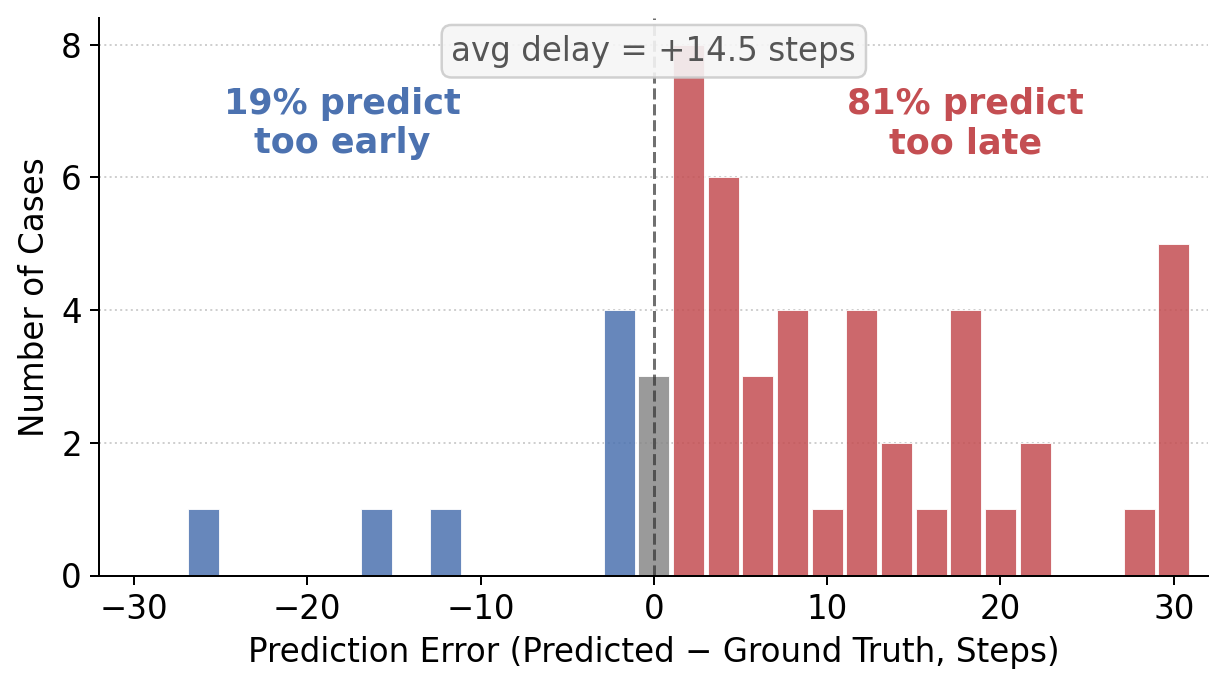}
    \caption{Distribution of prediction offsets (predicted step - ground-truth step) among mislocalized cases.}
    \label{fig:error_dist}
\end{figure}

To better understand these failures, we identify two recurring patterns among TrajAudit's mislocalized cases. (1) Some agents commit to a flawed problem framing in the initial planning stage before making any concrete code change. TrajAudit is more reliable when erroneous reasoning is reflected in observable code or tool-use evidence, but less effective when the decisive error remains implicit in an early plan. Consequently, it often selects a later step where the flawed plan first becomes externally observable. (2) Even when the root cause appears in code early in the trajectory, its effects may accumulate silently over many subsequent steps, leading TrajAudit to identify the step where the effects of the error become most evident rather than where the mistake was first introduced. Both patterns produce the same directional bias: among mislocalized predictions, 81\% are later than the annotated step, with an average delay of 14.5 steps, as illustrated in Figure~\ref{fig:error_dist}.

Successful localizations typically contain a distinct transition from correct exploration to one identifiable erroneous decision that directly causes the task failure. These findings suggest that future work should strengthen reasoning-level localization, especially for early planning errors whose consequences only become observable much later in the trajectory.
\section{Threats to Validity}
\textbf{Internal validity.} The main internal threats concern annotation reliability and the nondeterminism of LLM-based methods. We mitigate annotation subjectivity by using a unified guideline, independent annotation by two annotators, and arbitration by a third validator. For LLM-based methods and evaluators, we set the temperature to 0 across all experiments to reduce sampling variance.

\textbf{External validity.} The main external threat concerns generalizability. First, RootSE contains 102 instances filtered from 500 trajectories, excluding cases primarily attributable to external factors such as ambiguous task descriptions. This filtering improves annotation reliability and aligns with our focus on agent reasoning and action failures, but may bias the dataset away from failures caused by task ambiguity or other external conditions. Second, RootSE currently covers bug fixing and feature implementation tasks, and does not include other repository-level coding tasks such as test case generation~\cite{10329992}, which may limit the generalizability of our findings. At the same time, RootSE is comparable in scale to related benchmarks, including Who\&When~\cite{zhang2025which} with 184 instances and the study by Bouzenia et al.~\cite{bouzenia2025understanding} with 120 trajectories. Future work will extend the benchmark to broader task types and failure sources.
\section{Conclusion}
In this paper, we present TrajAudit, an automated failure diagnosis framework for repository-level coding agentic trajectories. By combining semantic saliency folding with prior failure reasoning, TrajAudit filters irrelevant context and guides an investigator agent toward likely causal steps in noisy trajectories. We also construct RootSE, a benchmark of 102 failed repository-level coding task trajectories with step-level earliest-error annotations. Experimental results demonstrate that TrajAudit consistently outperforms all baselines on RootSE, demonstrating its effectiveness.

\section*{Data Availability}
The complete RootSE benchmark, the TrajAudit implementation, and all artifacts required to reproduce the reported results are publicly available at: \url{https://github.com/LogAnalysisTech/TrajAudit}.

\bibliographystyle{IEEEtran}
\bibliography{references} 

@inproceedings{zhang2024autocoderover,
author = {Zhang, Yuntong and Ruan, Haifeng and Fan, Zhiyu and Roychoudhury, Abhik},
title = {AutoCodeRover: Autonomous Program Improvement},
year = {2024},
isbn = {9798400706127},
publisher = {Association for Computing Machinery},
address = {New York, NY, USA},
url = {https://doi.org/10.1145/3650212.3680384},
doi = {10.1145/3650212.3680384},
booktitle = {Proceedings of the 33rd ACM SIGSOFT International Symposium on Software Testing and Analysis},
pages = {1592–1604},
numpages = {13},
location = {Vienna, Austria},
series = {ISSTA 2024}
}

@inproceedings{wang2024openhands,
  title={OpenHands: An Open Platform for AI Software Developers as Generalist Agents},
  author={Xingyao Wang and Boxuan Li and Yufan Song and Frank F. Xu and Xiangru Tang and Mingchen Zhuge and Jiayi Pan and Yueqi Song and Bowen Li and Jaskirat Singh and Hoang H. Tran and Fuqiang Li and Ren Ma and Mingzhang Zheng and Bill Qian and Yanjun Shao and Niklas Muennighoff and Yizhe Zhang and Binyuan Hui and Junyang Lin and Robert Brennan and Hao Peng and Heng Ji and Graham Neubig},
  booktitle={International Conference on Learning Representations},
  year={2024},
  url={https://api.semanticscholar.org/CorpusID:271404773}
}

@inproceedings{yang2024swe,
author = {Yang, John and Jimenez, Carlos E. and Wettig, Alexander and Lieret, Kilian and Yao, Shunyu and Narasimhan, Karthik and Press, Ofir},
title = {SWE-agent: agent-computer interfaces enable automated software engineering},
year = {2024},
isbn = {9798331314385},
publisher = {Curran Associates Inc.},
address = {Red Hook, NY, USA},
booktitle = {Proceedings of the 38th International Conference on Neural Information Processing Systems},
articleno = {1601},
numpages = {125},
location = {Vancouver, BC, Canada},
series = {NIPS '24}
}

@inproceedings{zhou2025gsm,
author = {Zhou, Yang and Liu, Hongyi and Chen, Zhuoming and Tian, Yuandong and Chen, Beidi},
title = {GSM-$\infty$: how do your LLMs behave over infinitely increasing reasoning complexity and context length?},
year = {2025},
publisher = {JMLR.org},
booktitle = {Proceedings of the 42nd International Conference on Machine Learning},
articleno = {3178},
numpages = {51},
location = {Vancouver, Canada},
series = {ICML'25}
}

@inproceedings{zhang2025which,
author = {Zhang, Shaokun and Yin, Ming and Zhang, Jieyu and Liu, Jiale and Han, Zhiguang and Zhang, Jingyang and Li, Beibin and Wang, Chi and Wang, Huazheng and Chen, Yiran and Wu, Qingyun},
title = {Which agent causes task failures and when? on automated failure attribution of LLM multi-agent systems},
year = {2025},
publisher = {JMLR.org},
booktitle = {Proceedings of the 42nd International Conference on Machine Learning},
articleno = {3076},
numpages = {17},
location = {Vancouver, Canada},
series = {ICML'25}
}

@inproceedings{bouzenia2025understanding,
author = {Bouzenia, Islem and Pradel, Michael},
title = {Understanding Software Engineering Agents: A Study of Thought-Action-Result Trajectories},
year = {2025},
publisher = {IEEE Press},
url = {https://doi.org/10.1109/ASE63991.2025.00234},
doi = {10.1109/ASE63991.2025.00234},
booktitle = {2025 40th IEEE/ACM International Conference on Automated Software Engineering (ASE)},
pages = {2846–2857},
numpages = {12},
location = {Seoul, Korea, Republic of}
}

@article{deng2025swe,
  title={Swe-bench pro: Can ai agents solve long-horizon software engineering tasks?},
  author={Deng, Xiang and Da, Jeff and Pan, Edwin and He, Yannis Yiming and Ide, Charles and Garg, Kanak and Lauffer, Niklas and Park, Andrew and Pasari, Nitin and Rane, Chetan and others},
  journal={arXiv preprint arXiv:2509.16941},
  year={2025}
}

@article{albrecht2018autonomous,
title = {Autonomous agents modelling other agents: A comprehensive survey and open problems},
journal = {Artificial Intelligence},
volume = {258},
pages = {66-95},
year = {2018},
issn = {0004-3702},
doi = {https://doi.org/10.1016/j.artint.2018.01.002},
url = {https://www.sciencedirect.com/science/article/pii/S0004370218300249},
author = {Stefano V. Albrecht and Peter Stone}
}

@inproceedings{franklin1996agent,
author = {Franklin, Stan and Graesser, Art},
title = {Is it an Agent, or Just a Program? A Taxonomy for Autonomous Agents},
year = {1996},
isbn = {3540625070},
publisher = {Springer-Verlag},
address = {Berlin, Heidelberg},
booktitle = {Proceedings of the Workshop on Intelligent Agents III, Agent Theories, Architectures, and Languages},
pages = {21–35},
numpages = {15},
series = {ECAI '96}
}

@article{luo2025large,
  title={Large language model agent: A survey on methodology, applications and challenges},
  author={Luo, Junyu and Zhang, Weizhi and Yuan, Ye and Zhao, Yusheng and Yang, Junwei and Gu, Yiyang and Wu, Bohan and Chen, Binqi and Qiao, Ziyue and Long, Qingqing and others},
  journal={arXiv preprint arXiv:2503.21460},
  year={2025}
}

@article{ge2025introducing,
  title={Who is introducing the failure? automatically attributing failures of multi-agent systems via spectrum analysis},
  author={Ge, Yu and Xie, Linna and Li, Zhong and Pei, Yu and Zhang, Tian},
  journal={arXiv preprint arXiv:2509.13782},
  year={2025}
}

@inproceedings{mialon2023gaia,
 author = {Mialon, Gr\'{e}goire and Fourrier, Cl\'{e}mentine and Wolf, Thomas and LeCun, Yann and Scialom, Thomas},
 booktitle = {International Conference on Learning Representations},
 editor = {B. Kim and Y. Yue and S. Chaudhuri and K. Fragkiadaki and M. Khan and Y. Sun},
 pages = {9025--9049},
 title = {GAIA: a benchmark for General AI Assistants},
 url = {https://proceedings.iclr.cc/paper_files/paper/2024/file/25ae35b5b1738d80f1f03a8713e405ec-Paper-Conference.pdf},
 volume = {2024},
 year = {2024}
}

@inproceedings{chaudhury2025epman,
    title = "{E}p{MAN}: Episodic Memory {A}ttentio{N} for Generalizing to Longer Contexts",
    author = "Chaudhury, Subhajit  and
      Das, Payel  and
      Swaminathan, Sarathkrishna  and
      Kollias, Georgios  and
      Nelson, Elliot  and
      Pahwa, Khushbu  and
      Pedapati, Tejaswini  and
      Melnyk, Igor  and
      Riemer, Matthew",
    editor = "Che, Wanxiang  and
      Nabende, Joyce  and
      Shutova, Ekaterina  and
      Pilehvar, Mohammad Taher",
    booktitle = "Proceedings of the 63rd Annual Meeting of the Association for Computational Linguistics (Volume 1: Long Papers)",
    month = jul,
    year = "2025",
    address = "Vienna, Austria",
    publisher = "Association for Computational Linguistics",
    url = "https://aclanthology.org/2025.acl-long.574/",
    doi = "10.18653/v1/2025.acl-long.574",
    pages = "11696--11708",
    ISBN = "979-8-89176-251-0"
}

@inproceedings{
pan2025multiagent,
title={Why Do Multiagent Systems Fail?},
author={Melissa Z Pan and Mert Cemri and Lakshya A Agrawal and Shuyi Yang and Bhavya Chopra and Rishabh Tiwari and Kurt Keutzer and Aditya Parameswaran and Kannan Ramchandran and Dan Klein and Joseph E. Gonzalez and Matei Zaharia and Ion Stoica},
booktitle={ICLR 2025 Workshop on Building Trust in Language Models and Applications},
year={2025},
url={https://openreview.net/forum?id=wM521FqPvI}
}

@inproceedings{
jimenez2023swe,
title={{SWE}-bench: Can Language Models Resolve Real-world Github Issues?},
author={Carlos E Jimenez and John Yang and Alexander Wettig and Shunyu Yao and Kexin Pei and Ofir Press and Karthik R Narasimhan},
booktitle={The Twelfth International Conference on Learning Representations},
year={2024},
url={https://openreview.net/forum?id=VTF8yNQM66}
}

@article{han2024llm,
  title={LLM multi-agent systems: Challenges and open problems},
  author={Han, Shanshan and Zhang, Qifan and Jin, Weizhao and Xu, Zhaozhuo},
  journal={arXiv preprint arXiv:2402.03578},
  year={2024}
}

@inproceedings{epperson2025interactive,
author = {Epperson, Will and Bansal, Gagan and Dibia, Victor C and Fourney, Adam and Gerrits, Jack and Zhu, Erkang (Eric) and Amershi, Saleema},
title = {Interactive Debugging and Steering of Multi-Agent AI Systems},
year = {2025},
isbn = {9798400713941},
publisher = {Association for Computing Machinery},
address = {New York, NY, USA},
url = {https://doi.org/10.1145/3706598.3713581},
doi = {10.1145/3706598.3713581},
booktitle = {Proceedings of the 2025 CHI Conference on Human Factors in Computing Systems},
articleno = {156},
numpages = {15},
location = {
},
series = {CHI '25}
}

@inproceedings{parnin2011automated,
author = {Parnin, Chris and Orso, Alessandro},
title = {Are automated debugging techniques actually helping programmers?},
year = {2011},
isbn = {9781450305624},
publisher = {Association for Computing Machinery},
address = {New York, NY, USA},
url = {https://doi.org/10.1145/2001420.2001445},
doi = {10.1145/2001420.2001445},
booktitle = {Proceedings of the 2011 International Symposium on Software Testing and Analysis},
pages = {199–209},
numpages = {11},
location = {Toronto, Ontario, Canada},
series = {ISSTA '11}
}

@inproceedings{
zhang2025agentracer,
title={AgenTracer: Who Is Inducing Failure in the {LLM} Agentic Systems?},
author={Guibin Zhang and Junhao Wang and Junjie Chen and Wangchunshu Zhou and Kun Wang and Shuicheng YAN},
booktitle={The Fourteenth International Conference on Learning Representations},
year={2026},
url={https://openreview.net/forum?id=l05DseqvuD}
}

@inproceedings{
hong2023metagpt,
title={Meta{GPT}: Meta Programming for A Multi-Agent Collaborative Framework},
author={Sirui Hong and Mingchen Zhuge and Jonathan Chen and Xiawu Zheng and Yuheng Cheng and Jinlin Wang and Ceyao Zhang and Zili Wang and Steven Ka Shing Yau and Zijuan Lin and Liyang Zhou and Chenyu Ran and Lingfeng Xiao and Chenglin Wu and J{\"u}rgen Schmidhuber},
booktitle={The Twelfth International Conference on Learning Representations},
year={2024},
url={https://openreview.net/forum?id=VtmBAGCN7o}
}

@misc{deshpande2025trail,
  title={TRAIL: Trace Reasoning and Agentic Issue Localization},
  author={Darshan Deshpande and Varun Gangal and Hersh Mehta and Jitin Krishnan and Anand Kannappan and Rebecca Qian},
  year={2025},
  eprint={2505.08638},
  archivePrefix={arXiv},
  primaryClass={cs.AI},
  url={https://arxiv.org/abs/2505.08638}
}

@inproceedings{tian2025selective,
author = {Tian, Yuan and Zhang, Tianyi},
title = {Selective prompt anchoring for code generation},
year = {2025},
publisher = {JMLR.org},
booktitle = {Proceedings of the 42nd International Conference on Machine Learning},
articleno = {2362},
numpages = {24},
location = {Vancouver, Canada},
series = {ICML'25}
}

@article{liu2024lost,
    title = "Lost in the Middle: How Language Models Use Long Contexts",
    author = "Liu, Nelson F.  and
      Lin, Kevin  and
      Hewitt, John  and
      Paranjape, Ashwin  and
      Bevilacqua, Michele  and
      Petroni, Fabio  and
      Liang, Percy",
    journal = "Transactions of the Association for Computational Linguistics",
    volume = "12",
    year = "2024",
    address = "Cambridge, MA",
    publisher = "MIT Press",
    url = "https://aclanthology.org/2024.tacl-1.9/",
    doi = "10.1162/tacl_a_00638",
    pages = "157--173"
}

@inproceedings{xia2023automated,
author = {Xia, Chunqiu Steven and Wei, Yuxiang and Zhang, Lingming},
title = {Automated Program Repair in the Era of Large Pre-Trained Language Models},
year = {2023},
isbn = {9781665457019},
publisher = {IEEE Press},
url = {https://doi.org/10.1109/ICSE48619.2023.00129},
doi = {10.1109/ICSE48619.2023.00129},
booktitle = {Proceedings of the 45th International Conference on Software Engineering},
pages = {1482–1494},
numpages = {13},
location = {Melbourne, Victoria, Australia},
series = {ICSE '23}
}

@inproceedings{wei2022chain,
author = {Wei, Jason and Wang, Xuezhi and Schuurmans, Dale and Bosma, Maarten and Ichter, Brian and Xia, Fei and Chi, Ed H. and Le, Quoc V. and Zhou, Denny},
title = {Chain-of-thought prompting elicits reasoning in large language models},
year = {2022},
isbn = {9781713871088},
publisher = {Curran Associates Inc.},
address = {Red Hook, NY, USA},
booktitle = {Proceedings of the 36th International Conference on Neural Information Processing Systems},
articleno = {1800},
numpages = {14},
location = {New Orleans, LA, USA},
series = {NIPS '22}
}

@article{schick2023toolformer,
  title={Toolformer: Language models can teach themselves to use tools},
  author={Schick, Timo and Dwivedi-Yu, Jane and Dess{\`\i}, Roberto and Raileanu, Roberta and Lomeli, Maria and Hambro, Eric and Zettlemoyer, Luke and Cancedda, Nicola and Scialom, Thomas},
  journal={Advances in neural information processing systems},
  volume={36},
  pages={68539--68551},
  year={2023}
}

@inproceedings{liurepobench,
  title={Repobench: Benchmarking repository-level code auto-completion systems},
  author={Liu, Tianyang and Xu, Canwen and McAuley, Julian},
  booktitle={International Conference on Learning Representations},
  volume={2024},
  pages={47832--47850},
  year={2024}
}

@inproceedings{weiser1984program,
author = {Weiser, Mark},
title = {Program slicing},
year = {1981},
isbn = {0897911466},
publisher = {IEEE Press},
booktitle = {Proceedings of the 5th International Conference on Software Engineering},
pages = {439–449},
numpages = {11},
location = {San Diego, California, USA},
series = {ICSE '81}
}

@INPROCEEDINGS{barrak2025traceability,
  author={Barrak, Amine},
  booktitle={2025 40th IEEE/ACM International Conference on Automated Software Engineering Workshops (ASEW)}, 
  title={Traceability and Accountability in Role-Specialized Multi-Agent LLM Pipelines}, 
  year={2025},
  volume={},
  number={},
  pages={315-322},
  doi={10.1109/ASEW67777.2025.00064}}

@inproceedings{hu2025compileagent,
  author       = {Li Hu and
                  Guoqiang Chen and
                  Xiuwei Shang and
                  Shaoyin Cheng and
                  Benlong Wu and
                  LiGangyang LiGangyang and
                  Xu Zhu and
                  Weiming Zhang and
                  Nenghai Yu},
  editor       = {Wanxiang Che and
                  Joyce Nabende and
                  Ekaterina Shutova and
                  Mohammad Taher Pilehvar},
  title        = {CompileAgent: Automated Real-World Repo-Level Compilation with Tool-Integrated
                  LLM-based Agent System},
  booktitle    = {Proceedings of the 63rd Annual Meeting of the Association for Computational
                  Linguistics (Volume 1: Long Papers), {ACL} 2025, Vienna, Austria,
                  July 27 - August 1, 2025},
  pages        = {2078--2091},
  publisher    = {Association for Computational Linguistics},
  year         = {2025},
  url          = {https://doi.org/10.18653/v1/2025.acl-long.103},
  doi          = {10.18653/V1/2025.ACL-LONG.103},
  bibsource    = {dblp computer science bibliography, https://dblp.org}
}

@inproceedings{shi2023large,
author = {Shi, Freda and Chen, Xinyun and Misra, Kanishka and Scales, Nathan and Dohan, David and Chi, Ed and Sch\"{a}rli, Nathanael and Zhou, Denny},
title = {Large language models can be easily distracted by irrelevant context},
year = {2023},
publisher = {JMLR.org},
booktitle = {Proceedings of the 40th International Conference on Machine Learning},
articleno = {1291},
numpages = {18},
location = {Honolulu, Hawaii, USA},
series = {ICML'23}
}

@inproceedings{qin2023toolllm,
  author       = {Yujia Qin and
                  Shihao Liang and
                  Yining Ye and
                  Kunlun Zhu and
                  Lan Yan and
                  Yaxi Lu and
                  Yankai Lin and
                  Xin Cong and
                  Xiangru Tang and
                  Bill Qian and
                  Sihan Zhao and
                  Lauren Hong and
                  Runchu Tian and
                  Ruobing Xie and
                  Jie Zhou and
                  Mark Gerstein and
                  Dahai Li and
                  Zhiyuan Liu and
                  Maosong Sun},
  title        = {ToolLLM: Facilitating Large Language Models to Master 16000+ Real-world
                  APIs},
  booktitle    = {The Twelfth International Conference on Learning Representations,
                  {ICLR} 2024, Vienna, Austria, May 7-11, 2024},
  publisher    = {OpenReview.net},
  year         = {2024},
  url          = {https://openreview.net/forum?id=dHng2O0Jjr},
  bibsource    = {dblp computer science bibliography, https://dblp.org}
}

@inproceedings{qian2024chatdev,
    title = "{C}hat{D}ev: Communicative Agents for Software Development",
    author = "Qian, Chen  and
      Liu, Wei  and
      Liu, Hongzhang  and
      Chen, Nuo  and
      Dang, Yufan  and
      Li, Jiahao  and
      Yang, Cheng  and
      Chen, Weize  and
      Su, Yusheng  and
      Cong, Xin  and
      Xu, Juyuan  and
      Li, Dahai  and
      Liu, Zhiyuan  and
      Sun, Maosong",
    editor = "Ku, Lun-Wei  and
      Martins, Andre  and
      Srikumar, Vivek",
    booktitle = "Proceedings of the 62nd Annual Meeting of the Association for Computational Linguistics (Volume 1: Long Papers)",
    month = aug,
    year = "2024",
    address = "Bangkok, Thailand",
    publisher = "Association for Computational Linguistics",
    url = "https://aclanthology.org/2024.acl-long.810/",
    doi = "10.18653/v1/2024.acl-long.810",
    pages = "15174--15186"
}

@article{landis1977measurement,
  title={The measurement of observer agreement for categorical data},
  author={Landis, J Richard and Koch, Gary G},
  journal={biometrics},
  pages={159--174},
  year={1977},
  publisher={JSTOR}
}

@article{cohen1960coefficient,
  title={A coefficient of agreement for nominal scales},
  author={Cohen, Jacob},
  journal={Educational and psychological measurement},
  volume={20},
  number={1},
  pages={37--46},
  year={1960},
  publisher={Sage Publications Sage CA: Thousand Oaks, CA}
}

@article{hou2024large,
author = {Hou, Xinyi and Zhao, Yanjie and Liu, Yue and Yang, Zhou and Wang, Kailong and Li, Li and Luo, Xiapu and Lo, David and Grundy, John and Wang, Haoyu},
title = {Large Language Models for Software Engineering: A Systematic Literature Review},
year = {2024},
issue_date = {November 2024},
publisher = {Association for Computing Machinery},
address = {New York, NY, USA},
volume = {33},
number = {8},
issn = {1049-331X},
url = {https://doi.org/10.1145/3695988},
doi = {10.1145/3695988},
journal = {ACM Trans. Softw. Eng. Methodol.},
month = dec,
articleno = {220},
numpages = {79}
}

@article{trofimova2025openhandstrajs,
 title={OpenHands Trajectories with Qwen3-Coder-480B-A35B-Instruct},
 author={Trofimova, Maria and Shevtsov, Anton and Ibragim, Badertdinov and Pyaev, Konstantin and Karasik, Simon and Golubev, Alexander},
 year={2025},
 journal={Nebius blog},
 note={}
}

@article{myers1986nd,
author = {Myers, Eugene W.},
title = {AnO(ND) difference algorithm and its variations},
year = {1986},
issue_date = {Nov 1986},
publisher = {Springer-Verlag},
address = {Berlin, Heidelberg},
volume = {1},
number = {1–4},
issn = {0178-4617},
url = {https://doi.org/10.1007/BF01840446},
doi = {10.1007/BF01840446},
journal = {Algorithmica},
month = nov,
pages = {251–266},
numpages = {16}
}

@inproceedings{lu2025exploring,
author = {Lu, Ruofan and Li, Yichen and Huo, Yintong},
year = {2025},
month = {11},
pages = {3856-3860},
title = {Exploring Autonomous Agents: A Closer Look at Why They Fail When Completing Tasks},
doi = {10.1109/ASE63991.2025.00330}
}

@inproceedings{du2017deeplog,
author = {Du, Min and Li, Feifei and Zheng, Guineng and Srikumar, Vivek},
title = {DeepLog: Anomaly Detection and Diagnosis from System Logs through Deep Learning},
year = {2017},
isbn = {9781450349468},
publisher = {Association for Computing Machinery},
address = {New York, NY, USA},
url = {https://doi.org/10.1145/3133956.3134015},
doi = {10.1145/3133956.3134015},
booktitle = {Proceedings of the 2017 ACM SIGSAC Conference on Computer and Communications Security},
pages = {1285–1298},
numpages = {14},
location = {Dallas, Texas, USA},
series = {CCS '17}
}

@INPROCEEDINGS{guo2021logbert,
  author={Guo, Haixuan and Yuan, Shuhan and Wu, Xintao},
  booktitle={2021 International Joint Conference on Neural Networks (IJCNN)}, 
  title={LogBERT: Log Anomaly Detection via BERT}, 
  year={2021},
  volume={},
  number={},
  pages={1-8},
  doi={10.1109/IJCNN52387.2021.9534113}}

@INPROCEEDINGS{he2016experience,
  author={He, Shilin and Zhu, Jieming and He, Pinjia and Lyu, Michael R.},
  booktitle={2016 IEEE 27th International Symposium on Software Reliability Engineering (ISSRE)}, 
  title={Experience Report: System Log Analysis for Anomaly Detection}, 
  year={2016},
  volume={},
  number={},
  pages={207-218},
  doi={10.1109/ISSRE.2016.21}}

@article{landauer2023deep,
title = {Deep learning for anomaly detection in log data: A survey},
journal = {Machine Learning with Applications},
volume = {12},
pages = {100470},
year = {2023},
issn = {2666-8270},
doi = {https://doi.org/10.1016/j.mlwa.2023.100470},
url = {https://www.sciencedirect.com/science/article/pii/S2666827023000233},
author = {Max Landauer and Sebastian Onder and Florian Skopik and Markus Wurzenberger}
}

@article{xia2025live,
  title={Live-SWE-agent: Can Software Engineering Agents Self-Evolve on the Fly?},
  author={Xia, Chunqiu Steven and Wang, Zhe and Yang, Yan and Wei, Yuxiang and Zhang, Lingming},
  journal={arXiv preprint arXiv:2511.13646},
  year={2025}
}

@inproceedings{
badertdinov2026swe,
title={{SWE}-rebench: An Automated Pipeline for Task Collection and Decontaminated Evaluation of Software Engineering Agents},
author={Ibragim Badertdinov and Alexander Golubev and Maksim Nekrashevich and Anton Shevtsov and Simon Karasik and Andrei Andriushchenko and Maria Trofimova and Daria Litvintseva and Boris Yangel},
booktitle={The Thirty-ninth Annual Conference on Neural Information Processing Systems Datasets and Benchmarks Track},
year={2026},
url={https://openreview.net/forum?id=nMpJoVmRy1}
}

@article{10.1016/j.infsof.2008.09.009,
author = {Kitchenham, Barbara and Pearl Brereton, O. and Budgen, David and Turner, Mark and Bailey, John and Linkman, Stephen},
title = {Systematic literature reviews in software engineering - A systematic literature review},
year = {2009},
issue_date = {January, 2009},
publisher = {Butterworth-Heinemann},
address = {USA},
volume = {51},
number = {1},
issn = {0950-5849},
url = {https://doi.org/10.1016/j.infsof.2008.09.009},
doi = {10.1016/j.infsof.2008.09.009},
journal = {Inf. Softw. Technol.},
month = jan,
pages = {7–15},
numpages = {9}
}

@article{10.1109/TSE.2014.2372785,
author = {Barr, Earl T. and Harman, Mark and McMinn, Phil and Shahbaz, Muzammil and Shin Yoo},
title = {The Oracle Problem in Software Testing: A Survey},
year = {2015},
issue_date = {May 2015},
publisher = {IEEE Press},
volume = {41},
number = {5},
issn = {0098-5589},
url = {https://doi.org/10.1109/TSE.2014.2372785},
doi = {10.1109/TSE.2014.2372785},
journal = {IEEE Trans. Softw. Eng.},
month = may,
pages = {507–525},
numpages = {19}
}

@inproceedings{318773.318946,
author = {Zeller, Andreas},
title = {Yesterday, my program worked. Today, it does not. Why?},
year = {1999},
isbn = {3540665382},
publisher = {Springer-Verlag},
address = {Berlin, Heidelberg},
booktitle = {Proceedings of the 7th European Software Engineering Conference Held Jointly with the 7th ACM SIGSOFT International Symposium on Foundations of Software Engineering},
pages = {253–267},
numpages = {15},
location = {Toulouse, France},
series = {ESEC/FSE-7}
}

@ARTICLE{10329992,
  author={Schäfer, Max and Nadi, Sarah and Eghbali, Aryaz and Tip, Frank},
  journal={IEEE Transactions on Software Engineering}, 
  title={An Empirical Evaluation of Using Large Language Models for Automated Unit Test Generation}, 
  year={2024},
  volume={50},
  number={1},
  pages={85-105},
  doi={10.1109/TSE.2023.3334955}}

@article{10.1145/3631974,
author = {Zhang, Quanjun and Fang, Chunrong and Ma, Yuxiang and Sun, Weisong and Chen, Zhenyu},
title = {A Survey of Learning-based Automated Program Repair},
year = {2023},
issue_date = {February 2024},
publisher = {Association for Computing Machinery},
address = {New York, NY, USA},
volume = {33},
number = {2},
issn = {1049-331X},
url = {https://doi.org/10.1145/3631974},
doi = {10.1145/3631974},
journal = {ACM Trans. Softw. Eng. Methodol.},
month = dec,
articleno = {55},
numpages = {69}
}

\end{document}